\documentclass[journal,twocolumn]{IEEEtran}
\usepackage{diagbox} % 加载宏包
\usepackage{graphicx}
\usepackage{epstopdf}
\usepackage{psfrag}
\usepackage{subfigure}
\usepackage{url}
\usepackage{stfloats}
\usepackage{amsfonts,amssymb,amsmath,bm,paralist,theorem,cite,ifthen,color,nccmath}

\usepackage{calc}
\usepackage{enumerate}
\usepackage{multirow}
\usepackage{makecell}
\usepackage[ruled]{algorithm2e}
\usepackage{setspace}
\usepackage{array}
\usepackage{float}
\usepackage{soul}

\usepackage[colorlinks,
            linkcolor=red,
            anchorcolor=red,
            citecolor=red]{hyperref}
\usepackage[]{mdframed}

 \usepackage[figurename=Fig., labelsep=period, font=small]{caption}
\SetKwInOut{Input}{Input}
\SetKwInOut{Output}{Output}

\newcommand{\T}{{\scriptscriptstyle\mathsf{T}}}
\renewcommand{\H}{{\scriptscriptstyle\mathsf{H}}}

\usepackage[normalem]{ulem}

\usepackage{pifont}
\newcommand{\xmark}{\ding{55}} 

\graphicspath{{Figures/}}

\newcounter{algoline}

\AtBeginEnvironment{algorithm}{\setcounter{algoline}{0}}
\newcommand\Ccl{\ensuremath{\mathcal{C}}}
\newcommand\Dcl{\ensuremath{\mathcal{D}}}

\newcommand\Ncl{\ensuremath{\mathcal{N}}}

\newcommand\Icl{\ensuremath{\mathcal{I}}}

\newcommand\Tcl{\ensuremath{\mathcal{T}}}

\newcommand\Vcl{\ensuremath{\mathcal{V}}}

\newcommand\Xcl{\ensuremath{\mathcal{X}}}
\newcommand\Zcl{\ensuremath{\mathcal{Z}}}
\newcommand\Lcl{\ensuremath{\mathcal{L}}}

\newcommand\Cs{\ensuremath{{\mathbb{C}}}}
\newcommand\Es{\ensuremath{{\mathbb{E}}}}
\newcommand\Rs{\ensuremath{{\mathbb{R}}}}

\newcommand\Pbb{\ensuremath{{\mathbb{P}}}}

\newcommand\Ab{\ensuremath{ \mathbf{A} }}

\newcommand\Pb{\ensuremath{ \mathbf{P} }}

\newcommand\Xb{\ensuremath{ \mathbf{X} }}

\newcommand\Zb{\ensuremath{ \mathbf{Z} }}

\newcommand\bb{\ensuremath{ \mathbf{b} }}

\newcommand\fb{\ensuremath{ \mathbf{f} }}
\newcommand\hb{\ensuremath{ \mathbf{h} }}

\newcommand\pb{\ensuremath{ \mathbf{p} }}

\newcommand\vb{\ensuremath{ \mathbf{v} }}

\newcommand\xb{\ensuremath{ \mathbf{x} }}

\newcommand\zb{\ensuremath{ \mathbf{z} }}

\usepackage{relsize}
 \usepackage{afterpage}
\usepackage{etoolbox}
\newcommand{\zerodisplayskips}{%
  \setlength{\abovedisplayskip}{4pt}%
  \setlength{\belowdisplayskip}{4pt}%
  \setlength{\abovedisplayshortskip}{4pt}%
  \setlength{\belowdisplayshortskip}{4pt}}
\appto{\normalsize}{\zerodisplayskips}
\appto{\small}{\zerodisplayskips}
\appto{\footnotesize}{\zerodisplayskips}

\usepackage{titlesec}
\titlespacing{\section}{-0.64 cm}{2pt}{2pt}
\titlespacing{\subsection}{0 cm}{2pt}{2pt}
\usepackage[normalem]{ulem}
\usepackage[all=normal,paragraphs=tight,floats=normal,mathspacing=normal,wordspacing=tight,charwidths=tight,mathdisplays=normal,leading=normal]{savetrees}
\usepackage{tikz}
\usetikzlibrary{positioning,arrows.meta,fit,calc}
\usepackage{placeins}
\IEEEoverridecommandlockouts

\begin{document}

% Titles are generally capitalized except foTherefore, thechgain an, and, as,
% at, but, by, for, in, nor, of, on, or, the, to and avaiablityre usually
% Linebreaks \\ can be used within to get better formatdigital-to-analog converters DACs, or special symbols in the title.
%\trequiresitch-based Hybrid Beamforming for Wideband MIMO-OFDM Systems with Beam Squint}
\title{Knowledge Distillation for Sensing-Assisted Long-Term Beam Tracking in mmWave Communications}
% \title{Efficient Learning Design for Sensing-Assisted Beam Tracking in mmWave Communications}

\author{Mengyuan Ma, \IEEEmembership{Member, IEEE},
Nhan Thanh Nguyen, \IEEEmembership{Member, IEEE},
        % Ahmed Alkhateeb, \IEEEmembership{Member, IEEE},
        Nir Shlezinger, \IEEEmembership{Senior Member, IEEE},
        Yonina C. Eldar, \IEEEmembership{Fellow, IEEE},
        A.~Lee~Swindlehurst,~\IEEEmembership{Life Fellow,~IEEE},
        and Markku Juntti, \IEEEmembership{Fellow, IEEE}
        \thanks{This work was supported by the Research Council of Finland through 6G Flagship Program (grant 369116) and projects DIRECTION (grant 354901), DYNAMICS (grant 24305016), and CHIST-ERA PASSIONATE (grant 359817), by Business Finland, Keysight, MediaTek, Siemens, Ekahau, and Verkotan via project 6GLearn, by the HORIZON-JU-SNS-2023 project INSTINCT (101139161), and by U.S. National Science Foundation grant CCF-2225575. The authors wish to acknowledge CSC -- IT Center for Science, Finland, for computational resources.\textit{(Corresponding author: Mengyuan Ma.)}}
		\thanks{ Mengyuan Ma, Nhan Thanh Nguyen, and Markku Juntti are with Centre for Wireless Communications, University of Oulu, P.O.Box 4500, FI-90014, Finland (e-mail: \{mengyuan.ma, nhan.nguyen, markku.juntti\}@oulu.fi). Nir Shlezinger is with School of ECE, Ben-Gurion University of the Negev, Beer-Sheva 84105, Israel (email: nirshl@bgu.ac.il). A.~L.~Swindlehurst is with the with the Dept. of Electrical Engineering \& Computer Science, University of California, Irvine, CA, USA (e-mail: swindle@uci.edu). Yonina C. Eldar is with Faculty of Math and CS, Weizmann Institute of Science, Rehovot 76100, Israel (e-mail: yonina.eldar@weizmann.ac.il).}
        \vspace{-5mm}
        }

\maketitle

\begin{abstract}

Infrastructure-mounted sensors can capture rich environmental information to enhance communications and facilitate beamforming in millimeter-wave systems. This work presents an efficient sensing-assisted long-term beam tracking framework that selects optimal beams from a codebook for current and multiple future time slots. We first design a large attention-enhanced neural network (NN) to fully exploit past visual observations for beam tracking.  A convolutional NN extracts compact image features, while gated recurrent units with attention capture the temporal dependencies within sequences. The large NN then acts as the teacher to guide the training of a lightweight student NN via knowledge distillation. The student requires shorter input sequences yet preserves long-term beam prediction ability. Numerical results demonstrate that the teacher achieves Top-5 accuracies exceeding $93\%$ for current and six future time slots, approaching state-of-the-art performance with a $90\%$ reduction of model parameters. The student closely matches the teacher's performance while reducing the number of model parameters by over $1670\%$ and cutting complexity by over $450\%$, despite operating with $60\%$ shorter input sequences. This improvement significantly enhances data efficiency, reduces latency, and reduces power consumption in sensing and processing. %Our results highlight the promise of KD-driven sensing-assisted beam tracking for enabling accurate, efficient, and practical deployment in next-generation mmWave systems.

\end{abstract}

\begin{IEEEkeywords}
Beam prediction, beam tracking , sensing, deep learning, knowledge distillation.
\end{IEEEkeywords}

% page as needed:sacrifice less than $30\%$ SE can improve more than $200\%$ EE compared to the full-precision ones. Furthermore,
% \fi
\IEEEpeerreviewmaketitle

\section{Introduction}\label{sec: intro}
Millimeter wave (mmWave) and terahertz (THz) communication combined with large-scale multiple-input multiple-output (MIMO) systems promise to achieve high data rates to meet the demand for emerging applications, such as vehicular networks, unmanned aerial vehicles, and augmented/virtual reality \cite{jiang2021road}. However, a large number of antennas is required to steer narrow focused beams toward the target users in order to mitigate the severe path loss and guarantee the desired quality of service. Accurate beam tracking and alignment are essential to maintain reliable links, especially in high-mobility scenarios where the rapid variation of the radio environment can cause tracking errors and wasted resources for frequent link reestablishment \cite{yi2024beam}. %For example, a misalignment of $18$ degrees reduces the link budget by around $17$ dB and decreases the maximum throughput by up to $6$ Gbps or breaks the link entirely \cite{nitsche2015steering}.
Nonetheless, conventional beam tracking methods are based on lengthy codebook scanning, and, thus, typically incur significant overhead, posing challenges for real-time scenarios \cite{imran2024environment}. % Achieving swift, accurate, and robust beam tracking in these contexts necessitates increasingly sophisticated and advanced techniques.

With the advancement of large-scale MIMO at high frequencies, integrated sensing and communications (ISAC) has emerged as a promising paradigm and brings new opportunities for more efficient beam tracking \cite{cheng2022integrated}. Environmental information from the target users' surroundings can be captured by various sensors and leveraged to facilitate communications by machine learning (ML) techniques. Such sensing-assisted data-driven methods can quickly adapt to environmental variations and were recently demonstrated to facilitate rapid and high-performance beamforming \cite{shlezinger2024artificial,ma2024model}. This motivates the harnessing of sensory data to reduce beam training overhead and enable highly mobile mmWave communications systems.

\subsection{Prior Work}\label{sec:prior_work}
Conventional beam tracking has been largely based on predefined codebooks \cite{xiao2016hierarchical,li2019explore, qi2020hierarchical}. %Specifically, the transmitter (Tx) and receiver (Rx) simultaneously scan the beam space based on their codebooks to determine the best beam pair that returns the strongest signal strength.  
The most accurate and straightforward beam training approach is an exhaustive search over all possible transmitter--receiver beam pair combinations. Alternatively, heuristic strategies such as hierarchical search \cite{xiao2016hierarchical,qi2020hierarchical}, two-stage search \cite{li2019explore}, and adaptive beamforming \cite{jayaprakasam2017robust,lim2019beam,liu2020robust,zhang2019codebook,Zhang2020beam,ma2021closed} have been explored. %based on channel state information (CSI)  or past beams \cite{liu2020robust}.  
However, both exhaustive and heuristic search-based beamforming designs can incur significant training overhead and latency, particularly in large-scale MIMO systems. To address this challenge, deep learning-based beam training methods have been proposed \cite{alkhateeb2018deep,Shen2021Design,liu2022model,Fozi2022Fast,Ma2023continuous,nguyen2023deep,liu2024multimodal,Mu2021integrated,Liu2022learning,cao2023deep}. In particular, deep learning plays a key role in exploiting sensing data captured by LiDAR \cite{Jiang2024LiDAR,Klautau2019LiDAR}, radar \cite{Demirhan2022radar, Luo2023millimeter}, cameras \cite{Yang2023Environment,imran2024environment,charan2022vision,Xu20203D}, and global positioning systems (GPSs) \cite{charan2022vision} for beam management in communications. Such approaches fall under the category of sensing-assisted communications, which is among the main use cases in ISAC.

Beyond single-modality sensing, the joint exploitation of multiple sensing modalities provides enhanced and more robust performance \cite{Cheng2024Itelligent,Yang2023Environment}. This approach is generally referred to as multimodal sensing-aided communications. Charan~{\it et al.} \cite{charan2022vision} explored the use of both vision and GPS data for beam prediction, yielding better performance than when exploiting only one of the two modalities. LiDAR, radar, GPS, and cameras have been considered for joint beam prediction \cite{cui2024sensing,Tariq2024deep,tian2023multimodal,shi2024multimodal,park2025resource, zhang2025multimodal} using recurrent neural networks (RNNs) that leverage temporal dependencies \cite{shi2024multimodal} and for fusing data from distinct modalities using Transformers \cite{cui2024sensing,Tariq2024deep,tian2023multimodal,park2025resource}. Moreover, to enhance cross-modal feature extraction, Zhu {\it et al.} \cite{zhu2025advancing} designed a cross-attention module and used a dynamic fusion technique to improve beam prediction accuracy from radar and vision data. %A mixture-of-experts learning framework, which has trainable weights for features extracted from various sensing modalities, including vision, GPS, and LiDAR, was developed in \cite{zhang2025multimodal}.
However, the models for effectively processing multimodal data inevitably result in high complexity algorithms, calling for efficient solutions. 

A leading method in the ML literature to obtain efficient lightweight neural networks is based on 
 knowledge distillation (KD)  \cite{hinton2015distilling}, which transfers the knowledge from a well-trained large ``teacher'' to a small ``student'' model while guaranteeing a desired level of performance, enabling a form of model compression~\cite{zhang2021compacting}.  
KD is a regularization technique for transferring learned knowledge between classifiers, helping models to refine their predictions and enhance generalization \cite{phuong2019towards,tang2020understanding,mobahi2020self}. This is particularly valuable when distilling knowledge from a large, accurate teacher model to a lightweight student for deployment on resource-limited devices \cite{buciluǎ2006model,ba2014deep,polino2018model}. By learning to match the teacher’s soft outputs, which convey nuanced information about class relationships and decision boundaries, the student can achieve competitive performance with significantly reduced computational cost. Moreover, even without a separate teacher, a model can leverage its own outputs or intermediate representations as internal guidance to improve learning, a process known as self-distillation or self-KD. %Both KD and self-KD contribute to building models that are more efficient, robust, and accurate, especially in scenarios with limited training data or severe class imbalances. 
In the context of beam tracking, \cite{park2025resource} employed KD to obtain a compact model using only radar data, while in ISAC, KD has been applied to transceiver design \cite{kong2021knowledge}, channel estimation and feedback \cite{tang2021knowledge,catak2022defensive,guo2022environment}, semantic communications \cite{liu2023knowledge}, user positioning \cite{al2024knowledge}, and remote sensing \cite{zhang2021learning,ni2022cross}.

Despite the advances discussed above, most of the research pertinent to beam tracking focuses on predicting only the current beam based on current and/or past sensory data. Such methods result in frequent inference operations at each time step causing high overhead for sensing and processing in terms of power consumption, latency, and signaling. Long-term beam prediction can alleviate this by jointly predicting multiple future time steps, but this idea has remained largely unexplored.  Existing work  has considered long-term beam prediction based on LiDAR \cite{Jiang2024LiDAR}, radar \cite{Luo2023millimeter} and vision data \cite{jiang2022computer}, achieving encouraging results. %However, the LiDAR is costly for widespread usage.
To facilitate learning from vision data, the convolutional neural network (CNN) model YOLOv4 \cite{bochkovskiy2020yolov4} for object detection was used to extract the coordinates of potential targets from raw images in \cite{jiang2022computer}. However, YOLOv4, which has approximately $64$M parameters, is a relatively heavy model for deployment on resource-limited devices. More efficient long-term beam tracking requires further exploration. While LiDAR sensing provides accurate depth cues for beam prediction, cameras are significantly lower-cost and more widely deployed together with communication infrastructure. This motivates vision-only beam tracking to enable scalable and practical beam management without specialized sensing hardware.

\subsection{Contributions }\label{sec:contributions}
Long-term beam prediction in mmWave vehicular channels is challenging due to the highly dynamic and non-stationary propagation environment, where user mobility and rapid geometry evolution cause fast variations in the channel–beam relationship. As the prediction horizon increases, the correlation between past observations and future beam states weakens, making accurate prediction increasingly difficult. Achieving reliable long-term prediction typically requires longer historical observation sequences to capture the UE motion patterns, which increases sensing cost, energy consumption, signaling overhead, and computational complexity. In this work, we address this challenge by leveraging KD to transfer long-term beam evolution knowledge from a high-capacity teacher model to a lightweight student model operating with shorter observation sequences. This design enables efficient long-term beam prediction while maintaining low complexity and sensing overhead.

Specifically, in this work, we develop a compact model that learns to implement long-term beam selection for current and multiple future time slots with the objective of maximizing the overall spectral efficiency. To this end, we cast the problem as an ML classification task, which we tackle in two stages. First, we initially ignore model complexity and sensing overhead, and design a sequence-to-sequence (Seq2Seq) neural network model that consists of dedicated CNNs and gated recurrent unit (GRU) networks with an additional attention mechanism. The designed CNNs together with a simple yet effective image preprocessing method can extract compact image features. Temporal dependencies across time slots are captured by the GRUs with attention, enabling robust long-term beam prediction.

In this paper, we use the complex pre-trained model to obtain a lightweight beam selection model. We identify KD as a practically suitable paradigm for this task since, as opposed to alternative model compression frameworks such as pruning and weight quantization~\cite{zhang2021compacting}, KD enables the compressed student model to notably deviate from how the teacher is structured. Furthermore, KD transfers the teacher’s learned probabilistic knowledge instead of relying on pseudo-labels generated by classical algorithms as in algorithm-supervised learning approaches~\cite{shastri2025algorithm, morad2026sgd}. We thus employ the complex model to train a lightweight student that accepts shorter input sequences, reducing the  model complexity and minimizing the operational cost associated with sensing and data processing. %Extensive simulation results on a real-world dataset validate the effectiveness and superiority of the proposed approach.

The specific contributions of the paper are as follows:
\begin{itemize}
\item We develop an efficient end-to-end learning framework for long-term vision-based beam tracking by integrating CNNs, GRUs, and a multi-head attention (MHA) mechanism. This Seq2Seq model can effectively predict both current and future beams based on past sensor observations.% In this design, the CNNs extract representative spatial features from raw visual inputs, while the GRUs and MHA modules effectively capture and enhance temporal dependencies in the sequence.

\item We design a lightweight student ML model with depthwise separable convolution \cite{howard2017mobilenets} and convolutional block attention \cite{woo2018cbam} techniques for long-term beam tracking. The efficiency of the student model is guaranteed by the KD technique. Self-KD is leveraged to train a high-performing teacher model before imparting the knowledge to the student.

\item We further enhance the student model to generate highly accurate beams with shorter input sequences. This significantly alleviates the computational burden, power consumption, and latency associated with continuous sensing data acquisition and processing.% This contributes to reduced power consumption, lower latency in sensing and processing, and minimized signaling overhead, thereby enhancing the practicality of the proposed approach for energy-efficient and low-latency deployments.

\item Finally, we perform extensive simulations based on a realistic dataset to demonstrate the proposed framework. The results show that the teacher model achieves over $93\%$ Top-5 beam prediction accuracies across both current and six future time slots, approaching state-of-the-art performance with a $90\%$ complexity reduction. Remarkably, the student model can perform similarly to the teacher despite employing $1670\%$ fewer parameters, $450\%$ less complexity, and $60\%$ shorter input sequences. %thereby improving both data efficiency and deployment flexibility.
\end{itemize}

The enablers of our contributions lie in the KD technique, advanced neural network structures, and efficient data preprocessing, which are judiciously designed for long-term beam tracking. Unlike \cite{jiang2022computer,Jiang2024LiDAR,Luo2023millimeter}, we design a dedicated CNN to extract representative semantic features from raw images and incorporate an attention mechanism to enhance the model's ability to capture temporal dependencies within the input sequence. Furthermore, we develop a lightweight student model and train it using the KD technique, enabling the use of fewer past RGB frames for beam tracking. Although KD was used in \cite{park2025resource} for developing a compact student model, the problem studied therein was the prediction of the current beam only, without considering the overhead for sensing and processing. In this work, KD is leveraged as a system-level enabler to transfer long-term beam evolution knowledge from a high-capacity teacher to a lightweight student model operating with shortened visual observation sequences. Beyond conventional KD usage for model compression, the proposed framework exploits KD to jointly improve computational efficiency and sensing/data efficiency, enabling low-latency and low-complexity vision-aided long-term beam tracking without sacrificing prediction accuracy. Although KD, sensing-assisted beam prediction, and sequence-based neural architectures have each appeared in prior works, jointly leveraging them to design and optimize a learning framework for both computational and data efficiency in long-term sensing-aided beam tracking remains underexplored. The specific differences between this paper and exiting works on sensing-aided beam tracking are summarized in Table~\ref{tb:Comparison of existing research}.
% considers the data efficiency of model input sequences with KD.

% \begin{table}[tb]
% \small
% % \renewcommand\arraystretch{1.5}
% \centering
% % \vspace{-0.2cm}
% % \vspace{-0.2cm}
% \caption{Comparison of this paper with prior works on sensing-aided beam tracking. \Nir{any chance the table can be made more visually aligned? it looks quite crammed in the title and awfully loose in its remainder...}}
% \label{tb:Comparison of existing research}
% % \begin{threeparttable}
%     \begin{tabular}{|p{1.5cm}|p{2.2cm}|p{1.9cm}|p{2cm}|} %l(left)居左显示 r(right)居右显示 c居中显示
%     \hline   
%      Reference(s)   &   \parbox{3cm}{Long-term\\Beam prediction} & \parbox{3cm}{Reduce model \\ complexity} &  \parbox{3cm}{Improve data \\efficiency}  \\
%       \hline
%       \hline       \cite{Klautau2019LiDAR,Demirhan2022radar, Luo2023millimeter,Yang2023Environment,imran2024environment,charan2022vision,Xu20203D,cui2024sensing,Tariq2024deep,tian2023multimodal,shi2024multimodal, zhang2025multimodal,zhu2025advancing}   & \xmark & \xmark& \xmark \\
%     \hline  
%      \cite{park2025resource}  & \xmark & \checkmark &\xmark \\%[5ex]
%     \hline 
%     \cite{jiang2022computer,Jiang2024LiDAR}   & \checkmark&  \xmark  &\xmark \\%[5ex]
%         \hline 
%     This work    & \checkmark &  \checkmark  &  \checkmark   \\%[5ex]
%     \hline 
%     \end{tabular}   
% \end{table}

\begin{table}[tb]
\small
\centering
\caption{Comparison of this paper with prior work on sensing-aided beam tracking.}
\label{tb:Comparison of existing research}
\begin{tabular}{|p{1.8cm}|c|c|c|}
\hline   
\textbf{Reference(s)} & 
\textbf{\begin{tabular}{@{}c@{}}Long-term\\Beam prediction\end{tabular}} & 
\textbf{\begin{tabular}{@{}c@{}}Reduce model\\complexity\end{tabular}} & 
\textbf{\begin{tabular}{@{}c@{}}Improve data\\efficiency\end{tabular}} \\
\hline\hline       

\begin{tabular}{@{}l@{}}\cite{Demirhan2022radar,Yang2023Environment,charan2022vision,Xu20203D}\\ \cite{Klautau2019LiDAR, cui2024sensing,Tariq2024deep,tian2023multimodal,shi2024multimodal}\\
\cite{imran2024environment,zhu2025advancing,zhang2025multimodal}\end{tabular}& \xmark & \xmark & \xmark \\
\hline  
\cite{park2025resource} & \xmark & \checkmark & \xmark \\
\hline 
\cite{jiang2022computer,Jiang2024LiDAR,Luo2023millimeter} & \checkmark & \xmark & \xmark \\
\hline 
\textbf{This work} & \checkmark & \checkmark & \checkmark \\
\hline 
\end{tabular}   
\end{table}

The rest of the paper is organized as follows. In Section~\ref{sec:system model}, we present the system model and problem formulation, and in Section~\ref{sec:vision-based beam tracking} we introduce the long-term vision-based beam-tracking design. We then delve into the KD-aided learning approach in Section~\ref{sec:KD aided learning}. Finally, we provide simulation results and conclusions in Sections~\ref{sec:simulation} and~\ref{sec:conclusion}, respectively. 

Throughout the paper, scalars, vectors, and matrices (or tensors) are denoted by lower\-case, boldface lowercase, and boldface uppercase letters, respectively. The expectation operation is represented by $\Es\left[\cdot\right]$. We use $\left|a \right|$ and $\left|\Ab \right|$ to denote the absolute value of $a$ and the matrix (tensor) containing the absolute value of the entries of $\Ab$, respectively.

\section{System Model and Problem formulation}\label{sec:system model}
% In this section, we first introduce the considered system model for mmWave communications. Then, we formulate beam tracking into an optimization problem. After that, we clearly define the vision-aided beam tracking machine learning task. Lastly, we also present a baseline beam tracking ML task using the previous optimal beam sequence.
\subsection{System Model}\label{sec:system model_A}

 \begin{figure}[t]
\vspace{-3mm}
	\small
		% \vspace{-4mm}
	\centering	
	\includegraphics[width=0.45\textwidth]{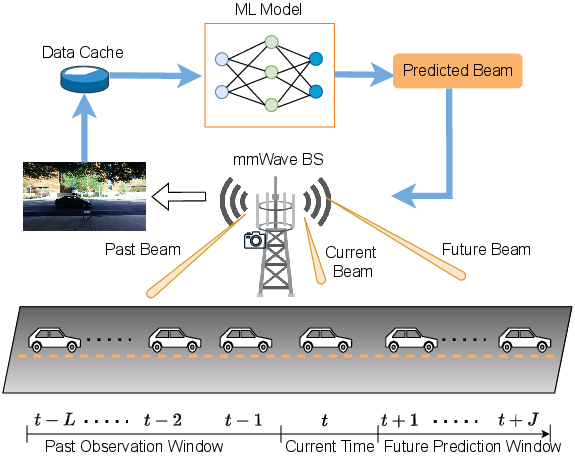}
	% \vspace{-2mm}
	\caption{Illustration of the considered system model. The BS senses the environment and the moving UE with an RGB camera. The sensory data are collected and cached for beam tracking using the designed ML model.}
	\label{fig:system model}
	% \vspace{-4mm}
\end{figure}
We consider a downlink mmWave system, where the base station (BS) serves a single-antenna mobile user equipment (UE). The BS is equipped with a uniform linear array of $N$ elements and an RGB camera (visual data sensor), as illustrated in Fig.~\ref{fig:system model}. We assume that the UE is visible to the sensing infrastructure and a line-of-sight (LoS) mmWave link exists\footnote{When the LoS path is blocked, distributed camera sensing can mitigate occlusions \cite{imran2024environment}, while proactive blockage prediction and multimodal sensing can further improve robustness \cite{charan2022computer,radarblockageprediction,wu2022lidar}.}. At time step $t$, the BS transmits $s[t]\in \Cs$ with normalized symbol power $\Es\left[|s|^2\right]=1$ to the UE. We assume a block fading channel between time slots.  Let  $\vb[t] \in \Cs^{N}$ denote the beamforming vector at time step $t$ with $\Es\left[\vb[t]^\H\vb[t]\right]=P$, where $P$ is the transmit power budget. Then, the received signal $y[t]$ is given as
\begin{align}\label{eq:signal model}
y[t] = \hb[t]^\H \vb[t] s[t] + n[t],
\end{align}
where $\hb[t] \in \Cs^{N }$ denotes the channel between the BS and the UE at time step $t$, and $n[t]\sim\Ccl\Ncl(0, \sigma_{\rm n}^2)$ is additive white Gaussian noise (AWGN) with power $\sigma_{\rm n}^2$. The signal-to-noise ratio (SNR) at time slot $t$ is thus given by \begin{align}
    \text{ SNR}[t]=\frac{|\hb[t]^\H \vb[t]\big|^2}{\sigma_{\rm n}^2}.
    \label{eq_SNR}
\end{align}

\subsection{Problem Formulation}\label{sec:system model_B}

At the current time slot $t$, the goal is to have the BS determine the beamformers for the current and $J$ future time slots, i.e., $\{t, t+1, \ldots, t+J\}$. The spectral efficiency over these $J+1$ time slots is
\begin{equation}
    R_J=\sum_{\tau=t}^{t+J}\log\left(1+ \text{ SNR}[\tau]\right).
\end{equation}
Let $\Vcl = \{\vb_1,\ldots, \vb_{|\Vcl|}\}$ and $\Icl_{\Vcl}=\{1,\ldots,|\Vcl|\}$ denote the beamforming codebook and  its associated index set. The beam tracking problem is expressed as
\begin{equation}\label{pb:P1}
    \underset{\vb[\tau] \in \Vcl, \forall \tau}{\rm maximize} \quad  R_J.
\end{equation}
For low SNR scenarios, we can approximate $R_J$ and reformulate problem \eqref{pb:P1} as \cite{Jiang2024LiDAR,imran2024environment,thomas2006elements}
\begin{equation}\label{pb:P2}
    \underset{\vb[\tau] \in \Vcl, \forall \tau}{\rm maximize} \quad  \sum_{\tau=t}^{t+J}|\hb[\tau]^\H \vb[\tau]\big|^2.
\end{equation}
Let $\bb^{\star}[t]=\big[b^{\star}[t],b^{\star}[t+1],\ldots,b^{\star}[t+J]\big]^\T$ represent the vector of beam indices corresponding to the optimal solution of \eqref{pb:P2}, i.e., $b^{\star}[t]= \arg\max_{b[t]\in \Icl_{\Vcl}} |\hb[t]^\H \vb_{b[t]}\big|^2$. Then problem \eqref{pb:P2} can be expressed as
\begin{equation}\label{pb:P3}
      \bb^{\star}[t] = \underset{b[\tau]\in \Icl_{\Vcl},\forall \tau}{\arg\max} \quad \sum_{\tau=t}^{t+J}|\hb[\tau]^\H \vb_{b[\tau]}\big|^2.
\end{equation}

The optimal solution to \eqref{pb:P3} can be obtained by decoupling it into $J+1$ subproblems with each solved via an exhaustive search over the $|\mathcal{V}|$ candidate beams. However, the complexity of such a method scales as $J|\mathcal{V}|$, which can incur high latency, especially with the large codebooks used in massive MIMO systems. Moreover, this approach requires perfect channel state information (CSI) at not only the current time slot, but also the \( J \) future time slots, which is generally unavailable in practice. 

In this work, we consider CSI-free beam tracking, where instead of aiming to recover $\bb[t]$ based on knowledge of $\hb[t]$, we utilize sensed visual data, denoted by $\Zb[t]$. Accordingly, our aim is to design a CSI-free mapping from $\Zb[t]$ into $\bb[t]$, such that the results remains effective with respect to the CSI-based performance measure in \eqref{pb:P3}. Unlike \cite{Demirhan2022radar, Yang2023Environment,imran2024environment,charan2022vision,Xu20203D,cui2024sensing,Tariq2024deep,tian2023multimodal,shi2024multimodal, zhang2025multimodal,zhu2025advancing,park2025resource} which address problem \eqref{pb:P3} by decoupling it into $J+1$ subproblems, we propose an efficient learning framework that directly solves problem \eqref{pb:P3} for long-term beam tracking, as will be elaborated below.

\section{Vision-based Long-term Beam Tracking}\label{sec:vision-based beam tracking}

\subsection{ML Task Definition}
%The vision sensory information of the mobile UE and the surrounding environment can be captured at the BS by its RGB camera.
Let $\Zb[t]\in\Rs^{3\times d_{\rm H}\times d_{\rm W}}$ denote the RGB image obtained at time slot $t$, where the dimension $3$ corresponds to the number of RGB channels, and $d_{\rm H}$ and $d_{\rm W}$ respectively represent the image height and width in pixels. Let $\Zcl[t] $ denote the sequence of sensory data, i.e., RGB images, from the $L$ previous time slots to the current time $t$, defined by $\Zcl[t]=\{\Zb[t-L],\Zb[t-L+1], \ldots,\Zb[t]]\}$. The objective of the learning task is to predict the optimal beams (equivalently the optimal beam indices in $\Icl_{\Vcl}$) for the current time slot $t$ and the $J$ future time slots $t+1,\ldots, t+J$. This problem can be cast as an ML classification task, where the number of classes $C$ is the size of the codebook, i.e., $C=\lvert \Vcl \rvert$. 

Denote the data preprocessing operations by $\Xcl[t]=g(\Zcl[t])$, mapping the input sequence to the ML model. Let $f(\Xcl[t];\Theta)$ denote the ML model with learnable parameters $\Theta$. The ML model outputs the probabilities of all possible beams at the $J+1$ (current and future) time slots. Let $p_c[t + j]$ denote the probability of selecting the $c$-th beam in the codebook at time slot $t + j$, and define $\pb[t + j] = [p_1[t + j], \ldots, p_{C}[t + j]]^\T \in \Rs^{C}, j=0,\ldots,J$. The intended output of the ML model is a probability distribution matrix given by

\begin{align}
    f(\Xcl[t];\Theta) = [\pb[t],\ldots,\pb[t+J]] \triangleq \Pb[t] \in \Rs^{C\times (J+1)}. \label{eq_output}
\end{align}
The predicted beam index is obtained as
\begin{equation}
    \hat{b}[\tau]=\arg\max_{c\in\Icl_{\Vcl}} \; p_c[\tau] ,\ \tau=t,\ldots,t+J.
\end{equation}
 The desired ML model for vision-aided beam tracking can be written as
\begin{equation}\label{pb:ML task}
    \Theta^{\star}=\arg\max_{\Theta} \quad \sum_{\tau=t}^{t+J} \Pbb\{\hat{b}[\tau]=b^{\star}[\tau]\},
\end{equation}
where $\Pbb\{\cdot\}$ denotes an event probability.  We note that $J$ and $L$ are hyperparameters, which are determined empirically.

\subsection{Data Preprocessing}
The raw images captured by the camera contain rich information about the environment surrounding the mobile UE. However, directly using these raw images as input to the ML model poses significant challenges for efficient learning, as irrelevant information for beam tracking acts as interference and increases the computational burden, motivating the need for effective preprocessing techniques. One approach is to extract bounding boxes of potential objects using additional CNN-based detectors, as demonstrated in \cite{jiang2022computer,imran2024environment}. However, this approach introduces significant computational complexity. For instance, the well-known YOLOv4 model contains approximately $6.4 \times 10^7$ parameters. 

To overcome the complexity challenge, we adopt background subtraction \cite{cui2024sensing} to remove background and irrelevant information without incurring additional computational overhead. Specifically, the data preprocessing operations $g(\cdot)$ consist of three steps:
\begin{itemize}
    \item \textbf{\textit{Step~1:}} First, we resize and transform the raw RGB images to grayscale. This step effectively diminishes the dimensions of the trainable tensors, reducing the complexity of the ML model. We denote
    \begin{align}
        \Zcl'[t]=\{\Zb'[t-L],\Zb'[t-L+1], \ldots, \Zb'[t]]\} \label{eq_post_process_Z}
    \end{align}
    as the sequence of post-processed images with $\Zb'[\tau] \in \Rs^{ d_{\rm H}'\times d_{\rm W}'}, \forall \tau$ ($d_{\rm H}'<d_{\rm H},d_{\rm W}'<d_{\rm W}$).
    
    \item \textbf{\textit{Step~2:}} Based on $\Zcl'[t]$ in \eqref{eq_post_process_Z},  we further construct
    \begin{equation}\label{eq_X_prime}
        \Xcl'[t]=\{\Xb'[t-L+1], \Xb'[t-L+2],\ldots,\Xb'[t]\}, 
    \end{equation}
    where $\Xb'[t-l]=\big| \Zb'[t-l]-\Zb'[t-l-1]\big|, l=0,\ldots,L-1$
    represents the {\it difference image} that highlights moving objects. This operation can effectively remove interference due to static objects, i.e., the background noise. However, useful information pertinent to the moving UE can also be compromised.
    \item \textbf{\textit{Step~3:}} To enhance the difference image, we construct a sequence of motion masks, defined as  
    \begin{align}
    \Xcl[t]=\{\Xb[t-L+1],\Xb[t-L+2], \ldots,\Xb[t]\}, \label{eq_X}
    \end{align}
    which are obtained by setting the large values in $\Xb'[\tau]$ above a given threshold equal to one, and setting all others to zero. Neglecting the time stamp $\tau$, the motion mask is given by:
\begin{equation}\label{eq:threolding}
   \Xb\left(m,n\right)= \begin{cases}
1 & {\rm  if } \; \Xb'(m,n) \geq \varepsilon \max(\Xb'), \\
 0 & {\rm otherwise},
\end{cases},
\end{equation}
where $\Xb(m,n)$ denotes the entry at the $m$-th row and $n$-th column of $\Xb$, and $\max(\Xb')$ returns the maximum value in $\Xb'$. Here, $\varepsilon\in(0,1)$ controls the percentage of information retained about the UE. A small $\varepsilon$ preserves most of this information and thus we set it to $0.1$ in the subsequent numerical experiments in Section~\ref{sec:simulation}. %Here, $\Xcl[t]$ in \eqref{eq_X} denotes the preprocessed data, which is ready to be fed into the beam tracking ML model.

\end{itemize}
We will illustrate the efficiency of this preprocessing procedure via simulation results in Section~\ref{sec:simulation}.

\subsection{ML Model Design}\label{sec:model design}
 \begin{figure*}[t]
% \vspace{-3mm}
	\small
		\vspace{-6mm}
	\centering	
	\includegraphics[width=0.9\textwidth]{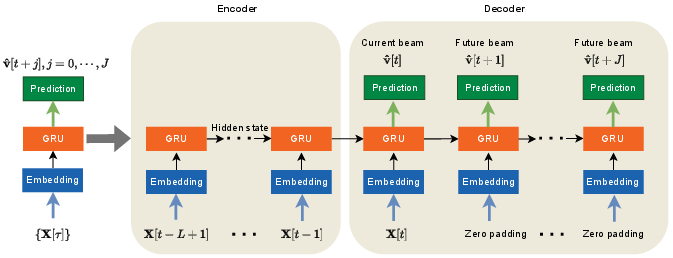}
	\vspace{-2mm}
	\caption{Illustration of the ML model.}
	\label{fig:ML model}
	% \vspace{-4mm}
\end{figure*}

 \begin{figure*}[t]
\vspace{-3mm}
	\small
		% \vspace{-4mm}
	\centering	
	\includegraphics[width=0.9\textwidth]{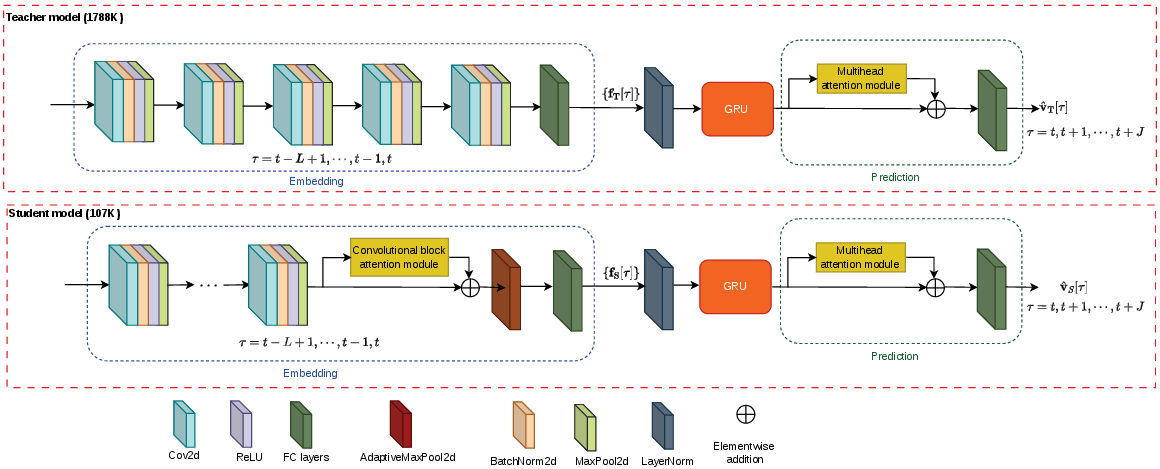}
	\vspace{-2mm}
	\caption{Illustration of the teacher and student model structures.}
	\label{fig:agent model}
	\vspace{-4mm}
\end{figure*}

\subsubsection{Motivation}
As seen in \eqref{pb:ML task}, the ML task is to predict a sequence of beam selections over current and future time steps based on a time series sequence of past images. Such Seq2Seq learning tasks have been widely studied in natural language processing (NLP) and computer vision. Well-known techniques such as RNNs~\cite{sherstinsky2020fundamentals} and attention-based mechanisms \cite{vaswani2017attention,achiam2023gpt,liu2024deepseek} have been developed for text and video generation. Transformer-based architectures \cite{vaswani2017attention,liu2020convtransformer,dosovitskiy2020image,yan2021contnet} have achieved competitive performance for image classification and large-language models (LLMs), and have been leveraged for multimodal sensing tasks \cite{park2025resource,zhu2025advancing,cui2024sensing}. Although Transformer-based ML models are powerful for capturing long-term dependencies between tokens, they typically involve substantially high computational complexity and memory footprint, which contradicts the lightweight and low-latency objective of this work.

Alternatively, we seek to design an efficient ML model employing RNN architectures and an attention mechanism. In particular, 
we consider using a GRU network \cite{dey2017gate}, which is a variant of the conventional RNN architecture. GRUs address the limitations of standard RNNs by incorporating gating mechanisms that regulate the flow of information, making them more effective at capturing long-term dependencies in sequential data. %Compared to plain vanilla RNNs, GRUs are better at mitigating the vanishing gradient problem, leading to more stable and efficient training. Moreover, compared with another RNN variant, known as LSTM, GRUs offer a favorable trade-off between model complexity and performance, often achieving better accuracy while requiring fewer parameters and less computational overhead \cite{chung2014empirical}. 

\subsubsection{Model Structures}\label{sec:model structures}

Fig.~\ref{fig:ML model} illustrates the considered GRU-based Seq2Seq model structure, which consists of three functional blocks: embedding, GRU, and prediction. The embedding block extracts features of the high-dimensional inputs $\Xb[t-l], l=0,\ldots,L-1$ to form low-dimensional feature vectors $\fb[t-l]\in \Rs^{D},l=0,\ldots,L-1$ in a latent space of size $D$. These feature vectors are then processed by the GRUs to obtain their sequential information, which is summarized by the output representation feature vectors $\fb'[t+j]\in \Rs^{D'},j=0,\ldots,J$. The prediction block further processes the representation features and outputs the predicted beam vector $\hat{\vb}[t+j], j=0,\ldots,J$. The entire process can be divided into two parts, i.e., the encoder and decoder. The encoder processes the input sequence $\{\Xb[\tau]\}$ (i.e., $\Xcl[t]$) step by step, updating its hidden state and ultimately producing a context vector that summarizes the information of the entire input. This context vector is then passed to the decoder, which generates the representation feature vectors $\fb'[t+j]\in \Rs^{D'},j=0,\ldots,J$ one by one, given the current data $\Xb[t]$. At each step, the decoder produces the current representation feature based on the previous hidden state. The hidden state in the GRUs helps the model efficiently capture temporal dependencies among the input features. %More details on GRU implementation and hidden states can be found in \cite{dey2017gate}.

\paragraph{Teacher Model}
To ensure strong embedding and prediction capability, we first design a high-capacity model, referred to as the teacher, and train a lightweight student model under its guidance via KD. The two models are illustrated in Fig.~\ref{fig:agent model}. Using pretrained backbones such as ResNet for image feature extraction \cite{cui2024sensing,Tariq2024deep,tian2023multimodal,shi2024multimodal,park2025resource} would incur high computational and memory costs. Instead, we develop a dedicated CNN-based embedding network tailored to the task. As summarized in Table~\ref{tab:teacher_arch}, where $I$ denotes the input sequence length, the teacher embedding block comprises five Conv–BN–ReLU–MaxPool layers, where Cov2d$(C_{\rm in},C_{\rm out})$ denotes a convolution with $C_{\rm in}$ input channels and $C_{\rm out}$ output channels, and BN denotes batch normalization. All convolution layers use $3\times3$ kernels and $2\times2$ max-pooling. The resulting feature map is projected to a compact $64$-dimensional representation via a fully connected embedding module. The feature sequence is then processed by layer normalization  and a two-layer GRU (hidden size $64$), followed by a multi-head attention (MHA) module of $8$ heads to capture temporal dependencies. Finally, a multi-layer perceptron (MLP) classifier produces beam probability outputs for multi-step prediction. The teacher model contains approximately $1.788$~M trainable parameters.

The MHA aims to provide a self-attention mechanism that allows each feature in a sequence to attend to all others and simultaneously learn different contextual information components. For the mathematical foundations of MHA, we refer the reader to \cite{vaswani2017attention}. Using MHA self-attention after the GRU combines the strengths of both sequential and attention-based modeling. Specifically, while GRUs effectively capture temporal dependencies in a step-by-step manner, they focus mainly on local dependencies. In contrast, MHA self-attention allows the model to directly attend to all positions in the sequence, enhancing its ability to capture global features. Furthermore, multiple attention heads enable the extraction of diverse features from the GRU outputs, leading to richer and more informative representations that can improve the model’s expressive ability and boost performance. The effectiveness of the MHA module will be justified using numerical experiments in Section~\ref{sec:simulation}.

\paragraph{Student Model}
Although the teacher model provides strong predictive capability, it suffers from high complexity and a large model size. The embedding block accounts for up to $95\%$ of the total parameters and dominates the overall computational cost. The student model aims to alleviate this burden while preserving effective feature extraction capability. To achieve a low complexity and memory footprint, we adopt depthwise separable convolution (DS-CNN) \cite{howard2017mobilenets} and a convolutional block attention (CBA) module \cite{woo2018cbam} to design the student embedding network.

DS-CNN factorizes a standard convolution into a depthwise and a pointwise ($1\times1$) convolution. The depthwise operation performs spatial filtering independently on each input channel, while the pointwise convolution linearly combines channel-wise features. This decomposition reduces multiply–accumulate operations by approximately a factor of $1/C_{\rm out}+1/k^2$ compared to conventional convolution with a $k\times k$ kernel size, enabling efficient spatial feature extraction with minimal accuracy loss.

To further enhance the representational power of the limited capacity student model, we incorporate a CBA attention mechanism \cite{woo2018cbam}. The CBA module sequentially infers channel attention and spatial attention maps to adaptively emphasize informative features while suppressing irrelevant responses. Channel attention captures inter-channel dependencies via global pooling and lightweight gating, whereas spatial attention highlights salient regions using aggregated feature statistics. By guiding the DS-CNN features toward task-relevant structures, CBA compensates for the reduced capacity of the lightweight embedding network and improves feature discriminability.

%  \begin{figure}[t]
% % \vspace{-3mm}
% 	\small
% 		% \vspace{-4mm}
% 	\centering	
% 	\includegraphics[width=0.5\textwidth]{Figs/Temporal_attention.eps}
% 	\vspace{-2mm}
% 	\caption{Temporal attention module.}
% 	\label{fig:temporal attention module}
% 	% \vspace{-4mm}
% \end{figure}
% \subsubsection{MHA Self-Attention Mechanism}

The detailed embedding architecture of the student model is summarized in Table~\ref{tab:student_arch}. The input image sequence is first processed by a shallow conventional CNN, followed by three DS-Conv blocks and a CBA module for efficient feature extraction. We use $3\times3$ convolution kernels for spatial feature extraction and $3\times3$ max-pooling for spatial downsampling. The resulting feature map is projected into a $64$-dimensional representation via a global pooling–based embedding module. The feature sequence is then processed by a single-layer GRU and an $8$-head MHA module to capture temporal dependencies, followed by an MLP classifier for beam prediction. As a result, the student model contains only $0.107$~M parameters, which is $16.7\times$ fewer than the teacher model.

\begin{table}[t]
\centering
\caption{The embedding architecture of the teacher model.}
\label{tab:teacher_arch}
\begin{tabular}{lc}
\hline
\textbf{Stage} & \textbf{ Module} \\
\hline
Input & Image sequence (I, 1, 224, 224) \\

\hline
\multicolumn{2}{l}{\textit{CNN Feature Extractor}} \\
Conv1 &  Conv2d(1, 4) $\to$ BN $\to$ ReLU $\to$ MaxPool \\
Conv2 &  Conv2d(4, 8) $\to$ BN $\to$ ReLU $\to$ MaxPool \\
Conv3 &  Conv2d(8, 16) $\to$ BN $\to$ ReLU $\to$ MaxPool \\
Conv4 &  Conv2d(16, 32) $\to$ BN $\to$ ReLU $\to$ MaxPool \\
Conv5 &  Conv2d(32, 64) $\to$ BN $\to$ ReLU $\to$ MaxPool \\
\hline
\multicolumn{2}{l}{\textit{Fully Connected Embedding}} \\
Flatten & Flatten \\
MLP & 3136$\to$512$\to$128$\to$64$\to$64\\

% \hline
% \multicolumn{2}{l}{\textit{Temporal Modeling}} \\
% LayerNorm & LN(64) \\
% GRU & 2-layer GRU (64 hidden) \\
% MHA & Multi-head attention  (NO. of head=$8$) \\
% \hline
% Classifier & MLP 64$\to$64$\to$64$\to$64 \\

\hline
\end{tabular}
\end{table}

\begin{table}[t]
\centering
\caption{The embedding architecture of the student model.}
\label{tab:student_arch}
\begin{tabular}{lc}
\hline
\textbf{Stage} & \textbf{ Module} \\
\hline
Input & Image sequence (I, 1, 224, 224)   \\

\hline
\multicolumn{2}{l}{\textit{Lightweight Feature Extractor}} \\
Conv1 & Conv2d(1, 18) $\to$ BN $\to$ ReLU $\to$ MaxPool \\
Block1 & DS-Conv(18, 36) $\to$ MaxPool   \\
Block2 & DS-Conv(36, 72) $\to$ MaxPool   \\
Block3 & DS-Conv(72, 144) $\to$ MaxPool   \\
CBA Module & Channel Attn $\to$ Spatial Attn \\
\hline
\multicolumn{2}{l}{\textit{Global Pooling Embedding}} \\
Global Pool & Adaptive MaxPool  \\
MLP &  144$\to$64$\to$64  \\

% \hline
% \multicolumn{2}{l}{\textit{Temporal Modeling}} \\
% LayerNorm & LN(64)  \\
% GRU & 1-layer GRU (64 hidden)  \\
% MHA & Multi-head attention (NO. of head=$8$)  \\
% \hline
% Classifier & MLP 64$\to$64$\to$64  \\
\hline
\end{tabular}
\end{table}

\section{Knowledge Distillation Aided Learning}\label{sec:KD aided learning}
As discussed in Section~\ref{sec:vision-based beam tracking}, the teacher model employs a more complex architecture than the student model and, thus can achieve superior performance in challenging beam prediction tasks. However, this performance gain comes at the cost of high computational complexity and a dependence on long input sequences spanning multiple time slots. These requirements lead to increased power consumption, higher latency, and greater overhead for data collection and preprocessing, posing significant challenges for practical deployments, especially in resource-constrained environments. 

While conventional model compression techniques such as pruning \cite{liu2018rethinking} or quantization \cite{rokh2023comprehensive} can reduce model complexity, they fail with shorter input sequences and thus are not useful for latency reduction. In contrast, KD provides both a compression mechanism and a learning framework, enabling the student model to inherit the predictive capability of the high-performing teacher model while being explicitly trained to function with shorter sequences. This makes KD particularly well-suited for our objective of achieving both low complexity and low-latency beam tracking.  We elaborate on the use of KD for efficient long-term beam tracking below.

\subsection{KD Loss Function}

The loss function plays a crucial role in KD-based training of the student model. In this section, we first present the details of the loss function employed to train the student model for beam tracking, given that the teacher model has already been developed as described in the previous section.

Let $\Lcl_{\rm task}$ and $\Lcl_{\rm distill}$ denote the task loss computed from the dataset and the distillation loss arising from the disparity between the teacher and student, respectively. In typical KD, the student minimizes a weighted combination of these two components:  
\begin{align}
    \label{eq:overall loss}
    \Lcl_{\rm s} = (1-\beta) \Lcl_{\rm task} + \beta \Lcl_{\rm distill}.
\end{align}
Here, a trade-off parameter $\beta \in [0,1]$ is employed to control the balance between these components: $\beta=0$ corresponds to learning purely from hard labels, whereas $\beta=1$ corresponds to learning solely from the teacher. While a large $\beta$ may lead the student to better mimic the teacher, it does not necessarily guarantee improved generalization. In fact, excessive reliance on the teacher can cause overfitting and degrade learning performance \cite{stanton2021does}. The value of $\beta$ is typically determined empirically. In the following, we formulate the specific loss components $\Lcl_{\rm task}$ and $\Lcl_{\rm distill}$ for the beam tracking task.

\subsubsection{Task Loss}
Let $\Dcl_{\rm}=\{\{\Zcl[t], \bb^\star [t]\}, t=0,\ldots, T\}$ denote the set of data sequences from the source dataset, where $T$ denotes the number of time slots over which vision data are collected, and $\Zcl[t]$ and $\bb^\star [t]$ are the input and label of the ML model, respectively. The received signal $y[t]$ in \eqref{eq:signal model} is leveraged to obtain $\bb^\star [t]$, which in mmWave communications is generally distributed non-uniformly among the $C$ candidate beams. Such a class imbalance among the datasets can lead to poor performance for the minority class. During training, we use the Focal loss \cite{lin2017focal} for $\Lcl_{\rm task}$, which is a modification of the standard cross-entropy loss designed to address the class imbalance problem. 

For the $\tau$-th sample $\Zb[\tau]$, the output of the ML model is the predicted probability vector $\pb [\tau]$ for the $C$ beam classes. Define the softmax function $\sigma(\xb)$ for a vector $\xb=[x_i,\ldots,x_N]$ as 
\begin{equation}
    \sigma_i(\xb)=\frac{\exp(x_i)}{\sum_{n=1}^N \exp(x_i) },
\end{equation}
and let $\zb[\tau]=[z_1[\tau],\ldots,z_C[\tau]] $ denote the vector of output logits. Then the $c$-th element of $\pb [\tau]$ is obtained as $p_c[\tau]=\sigma_c(\zb[\tau])$. The Focal loss for a single sample $\Zb[\tau]$ is given by
\begin{equation}\label{eq:Focal loss}
    l_{\rm Focal}[\tau]=-\alpha(1-p_{b^{\star}}[\tau])^{\gamma}\log\left(p_{b^{\star}}[\tau] \right),
\end{equation}
where $p_{b^{\star}}[\tau]$ denotes the predicted probability of selecting the ground-truth beam index $b^{\star}[\tau]$ at time slot $\tau$. The hyperparameter $\alpha$ is the weighting factor addressing class imbalance, and $\gamma$ is the focusing parameter that demphasizes easy examples. A large $\gamma$ leads to a small loss for well-classified samples, i.e., those with high output probabilities. This helps the model focus on difficult or misclassified samples, which are more informative. On the contrary, $\gamma=0$ leads to the conventional cross-entropy loss which treats all samples with equal importance. The overall task loss for the input sequence $\Zcl[t]$ is expressed as
\begin{equation}\label{eq:task loss}
    \Lcl_{\rm task}[t]=\sum_{\tau=t}^{t+J} l_{\rm Focal}[\tau]. 
\end{equation}

\subsubsection{Distillation Loss}
 We adopt the Kullback-Leibler (KL) divergence as the distillation loss $\Lcl_{\rm distill}$, which measures the similarity between the output distributions of the teacher and student models, respectively denoted as $\Tilde{P}_{\rm teacher}^{(\tau)}$ and $\Tilde{P}_{\rm student}^{(\tau)}$, given the input sample $\Zb[\tau]$. The KL divergence is computed as
\begin{equation}
    D_{\rm KL}\left( \Tilde{P}_{\rm teacher}^{(\tau)} \| \Tilde{P}_{\rm student}^{(\tau)} \right) = \sum_{c=1}^C \Tilde{P}_{\rm teacher}^{(\tau,c)} \log \left(\frac{\Tilde{P}_{\rm teacher}^{(\tau,c)}}{\Tilde{P}_{\rm student}^{(\tau,c)}} \right),
\end{equation}
where $\Tilde{P}_{\rm teacher}^{(\tau,c)}=\sigma_c(\zb_{\rm teacher}[\tau]/\Gamma)$, $\zb_{\rm teacher}[\tau]$ is the vector of output logits from the teacher model, and $\Tilde{P}_{\rm student}^{(\tau,c)}$ is similarly defined. Here, $\Gamma$ represents the temperature used to control the smoothness of the distribution; increasing $\Gamma $ makes the distribution more uniform, while $\Gamma \rightarrow 0$ results in the one-hot distribution. Therefore, an appropriate value for $\Gamma$ is needed to enable the student model to learn well from the teacher, and can be determined empirically. The distillation loss for the input sequence $\Zcl[t]$ is given by
\begin{equation}\label{eq: distillation loss}
    \Lcl_{\rm distill }[t]= \sum_{\tau=t}^{t+J}  D_{\rm KL}\left( \Tilde{P}_{\rm teacher}^{(\tau)} \| \Tilde{P}_{\rm student}^{(\tau)} \right) \cdot \Gamma^2,
\end{equation}
where the multiplication by $\Gamma^2$ arises since the gradients produced by the softmax function are scaled by $1/\Gamma$.

\subsection{Training Procedure}
Algorithm~\ref{alg1} summarizes the KD-aided learning procedure, where $f_{\rm T}(\cdot;\Theta_{\rm T})$ represents the pretrained teacher model; $\Dcl_{\rm tr}$ and $\Dcl_{\rm evl}$ denote the training and validation datasets obtained from the overall dataset $\Dcl$ without any overlap; $E$ represents the number of total epochs and $N_{\rm b}$ denotes the number of batches in each epoch. To begin the training, the model parameters are randomly initialized. For each epoch, $N_{\rm b}$ batches are generated by randomly dividing $\Dcl_{\rm tr}$ into batches of predefined size $B$. Steps 4--15 update the model parameters in a batch training manner. Data preprocessing is first performed in step 5, where $\Tcl^{(n)}=\{t_i^{(n)} \big| \Zcl[t_i^{(n)}] \in \Dcl^{(n)}_{\rm tr}, \forall i\}=\{t_1^{(n)},\ldots,t_B^{(n)}\}$ denotes the set of time stamps in the $n$-th batch. The $q$-th sequence sample $\{\Xcl[t_q^{(n)}]\}$ in the $n$-th batch is fed into the embedding block, which returns the low-dimensional feature vectors $\{\fb[t_q^{(n)}]\}$. After the GRU and prediction modules, the ML model outputs the predicted probability matrix $\Pb[t_q^{(n)}]$, which is leveraged to compute the sample task loss and distillation loss as indicated by steps 9 and 10, respectively. 

Next, the average task and distillation loss over a batch for the considered $J+1$ time slots are computed in steps 12 and 13. The overall loss is obtained and used to optimize the model parameters by the backpropagation algorithm in step 14. When batch learning is completed, the performance of the model $f_{\rm S}(\cdot;\Theta_{\rm S})$ is evaluated on the validation dataset $\Dcl_{\rm evl}$ in step 16. The optimal model parameters $\Theta^{\star}_{\rm S}$ are updated if a lower validation loss is found, as shown in steps 17--20. The optimal model parameters $\Theta^{\star}_{\rm S}$ are returned when the maximum number of epochs is reached or the best validation loss $L_{\rm evl}^{\star}$ stops decreasing over a predefined number of consecutive epochs.

\begin{algorithm}[t]
\small
% \setstretch{1}
\caption{KD-Aided Learning for Problem~\eqref{pb:ML task}.}\label{alg1}
\LinesNumbered %要求显示行号
% \LinesNumberedHidden
\KwIn{Training and validation datasets $\Dcl_{\rm tr}$, $\Dcl_{\rm evl}$; Pretrained teacher model $f_{\rm T}(\cdot;\Theta_{\rm T})$}
\KwOut{Student model parameters $\Theta_{\rm S}$}
% Perform greedy initialization:\\
 Initialize $\Theta^{\star}_{\rm S}$, $\Theta_{\rm S}=\Theta^{\star}_{\rm S}$, $ \Lcl_{\rm evl}^{\star}=1000$, $\beta$, $\Gamma$, and learning rate.\\
\For{$e=1,\ldots,E$}{
    Randomly divide $\Dcl_{\rm tr}$ into $N_{\rm b}$ batches $\{\Dcl^{(n)}_{\rm tr}\}_{n=1}^{N_{\rm b}}$ with batch size $B$. \\
    \For{$n=1,\ldots,N_{\rm b}$}{
        Perform data preprocessing $\Xcl[t]=g\left(\Zcl[t]\right)$ with $t\in \Tcl^{(n)}=\{t_1^{(n)},\ldots,t_B^{(n)}\}$ 
        
        \For{$q=1,\ldots,B$}{
        
            Obtain embedded feature $\{\fb[t_q^{(n)}]\}$ for $\{\Xcl[t_q^{(n)}]\}$. \\
            Feed  $\{\fb[t_q^{(n)}]\}$ into the GRU module with postprocessing of the prediction module. Compute the student model output $\Pb[t_q^{(n)}]=f_{\rm S}\left(\{\Xcl[t_q^{(n)}]\};\Theta_{\rm S}\right)$.\\

            Compute the task loss $\Lcl_{\rm task}[t_q^{(n)}]$ in \eqref{eq:task loss}.\\
            
            Compute the distillation loss $\Lcl_{\rm distill}[t_q^{(n)}]$ in \eqref{eq: distillation loss} based on $f_{\rm S}\left(\{\Xcl[t_q^{(n)}]\};\Theta_{\rm S}\right)$ and $f_{\rm T}(\{\Xcl[t_q^{(n)}]\};\Theta_{\rm T})$.

        }     
        Compute average task loss over the batch in $J+1$ time slots: $\Lcl_{\rm task}=\frac{1}{B(J+1)}\sum\limits_{ t\in \Tcl^{(n)}} \Lcl_{\rm task}[t]$.\\

        Compute average distillation loss over the batch in $J+1$ time slots: $\Lcl_{\rm distill}=\frac{1}{B(J+1)}\sum\limits_{ t\in \Tcl^{(n)}} \Lcl_{\rm distill}[t]$.

        Obtain the overall loss in \eqref{eq:overall loss} and update $\Theta_{\rm S}$ with an optimizer.
    }
    Compute validation loss $\Lcl_{\rm evl}^{(e)}$ based on $f_{\rm S}(\cdot;\Theta_{\rm S})$ and $\Dcl_{\rm evl}$.\\
    \If{$\Lcl_{\rm evl}^{(e)}< \Lcl_{\rm evl}^{\star}$}{
        update the best model $\Theta^{\star}_{\rm S}=\Theta_{\rm S}$. \\
        update the best loss $\Lcl_{\rm evl}^{\star}=\Lcl_{\rm evl}^{(e)}$.
    }
 }

    Return $\Theta^{\star}_{\rm S}$.
\end{algorithm}

To enhance the learning efficiency of the student model via KD, a good teacher is required. We first train the teacher model by setting $\beta=0$ in Algorithm~\ref{alg1}, and then refine it with self-KD. With the refined teacher model, we train the student model with KD. The effectiveness of the proposed KD-aided learning framework will be experimentally validated in the sequel.

\subsection{Signaling Protocol for Deployment}\label{sec:protocol}
During initial access, instead of performing exhaustive beam sweeping over the entire codebook, the proposed vision-aided framework first predicts a small set of candidate beams (e.g., Top-5) using the pretrained model. The gNB then performs beam sweeping only over these predicted candidates. The user measures the received signal power and feeds back the index corresponding to the strongest beam. Upon receiving this feedback, the gNB establishes the communication link using the selected beam and continues tracking the user based on subsequent Top-5 predictions generated by the deployed model. This procedure remains compatible with existing beam management protocols \cite{zecchin2022lidar,xue2024ai}. Moreover, since the model predicts beams over $J$ future time slots, sweeping and signaling updates can be performed less frequently, thereby reducing overhead and latency while maintaining reliable beam alignment.

\section{Numerical Results}\label{sec:simulation}
In this section, we provide extensive numerical simulations to demonstrate the performance of the proposed sensing-assisted long-term beam tracking approach\footnote{The simulation codes are available at \url{https://github.com/WillysMa/Sensing-Assisted-Beam-Tracking.git}.}. Experiments are based on Scenario 9 of the DeepSense 6G dataset \cite{alkhateeb2023deepsense}, which provides sensory data and optimal beams for real-world mmWave communications. 

\subsection{Dataset Preparation}

\begin{figure}[t]
% \vspace{-3mm}
	\small
		% \vspace{-4mm}
	\centering	
	\includegraphics[width=0.45\textwidth]{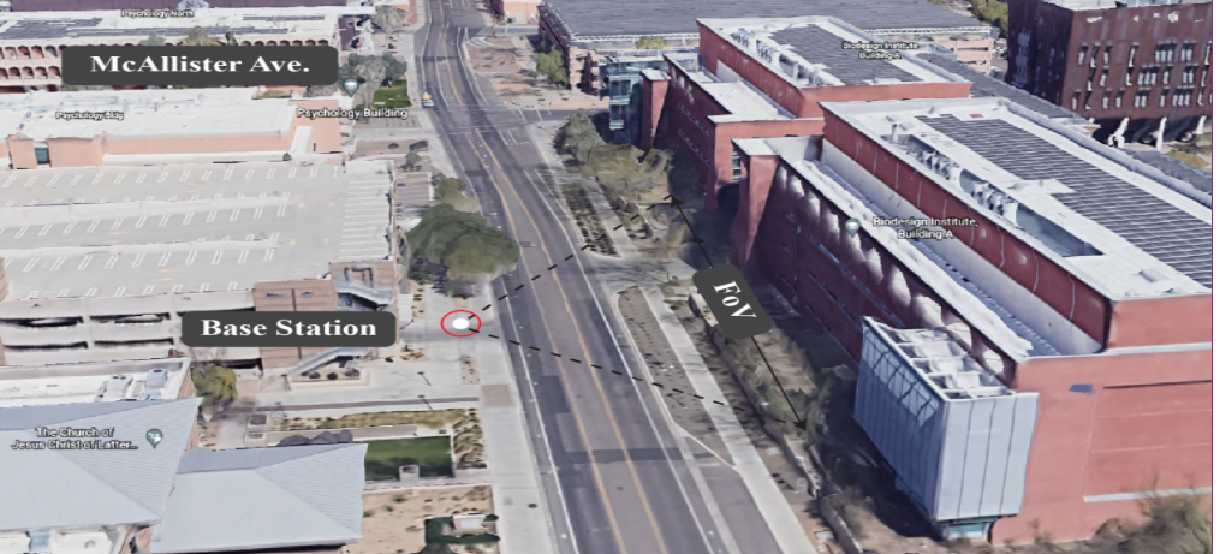}
	% \vspace{-2mm}
	\caption{Scenario 9 of the DeepSense 6G dataset.}
	\label{fig:scenario 9}
	% \vspace{-4mm}
\end{figure}

Fig.~\ref{fig:scenario 9} illustrates Scenario 9 from the DeepSense 6G dataset. A BS equipped with a 16-element ULA and an RGB camera is deployed roadside to receive signals from a moving UE, which transmits signals at $60$~GHz using a quasi-omni antenna. An oversampled beamforming codebook with 64 predefined beams is used at the BS. At each time step of duration of $128$ ms, the BS performs beam sweeping to measure the received power across all beams and captures an RGB image of the UE. The channel coherence time is approximately $143$ ms \cite{Jiang2024LiDAR}, which is longer than the sampling interval, justifying the assumption of a block fading channel in Section~\ref{sec:system model_A}. %This work focuses on a single-vehicle scenario; extensions to other UE types or multi-UE settings are left for future study.

\begin{figure}[t]
\vspace{-2mm}
	\small
		% \vspace{-4mm}
	\centering	
	\includegraphics[width=0.45\textwidth]{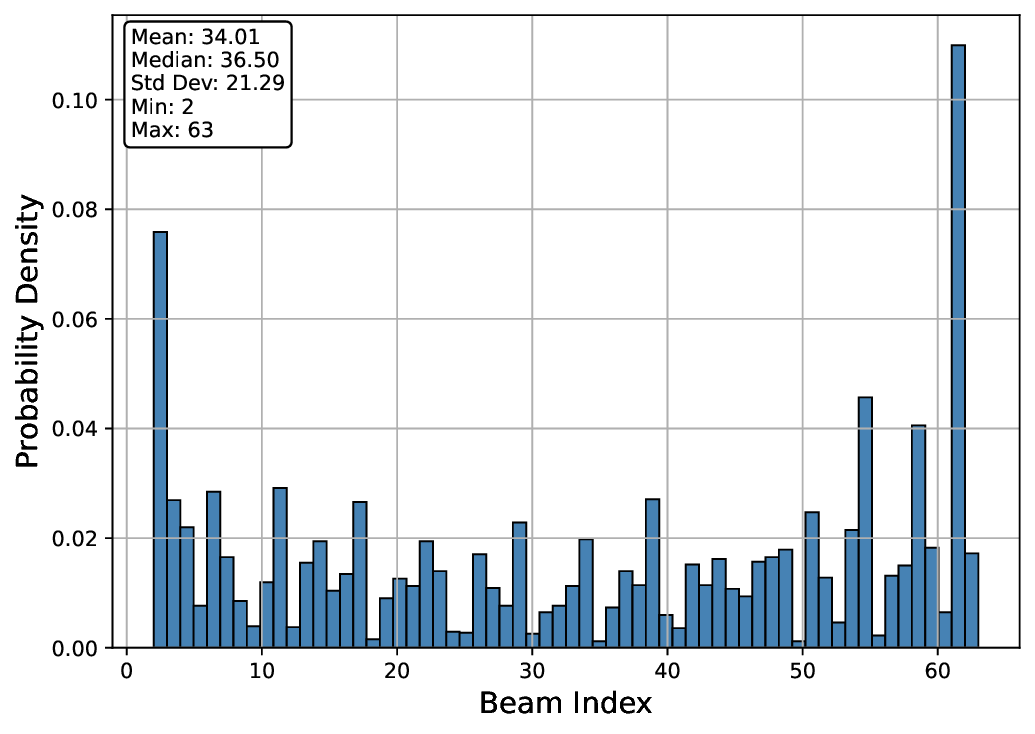}
	\vspace{-2mm}
	\caption{Statistics of the optimal beam index in the considered dataset.}
	\label{fig:statistics}
	% \vspace{-2mm}
\end{figure}

Scenario 9 of the DeepSense 6G dataset contains $5964$ samples that belong to multiple data sequences. In a data sequence, the moving UE passes by the BS, which acquires RGB images and the received power from each beam at multiple time steps. For any time step $t$, the maximum number of past images is set to $L=8$ in each sequence sample $\Zcl[t]$, while the number of future time steps for beam prediction is set to $J=6$. Therefore, the useful dataset contains a total of $T=4060$ samples $\{\Zcl[t], \bb^{\star}[t]\}$ to guarantee that the images and labels are within the same data sequence. The training and validation datasets consist of $80\%$ and $20\%$ of the total $4060$ samples, which corresponds to $3286$ and $774$ samples, respectively. Due to the limited data, the validation dataset is also used at the inference stage to evaluate the generalization performance of the ML models. Fig.~\ref{fig:statistics} shows the statistics of the dataset, demonstrating that the numbers of samples belonging to different classes are imbalanced. Therefore, it is reasonable to use the Focal loss in \eqref{eq:Focal loss} for training models. 

\subsection{Experiment Setup}\label{sec:experiment setup}
 \begin{figure*}[t]
% \vspace{-3mm}
	\small
		\vspace{-4mm}
	\centering	
	\includegraphics[width=0.8\textwidth]{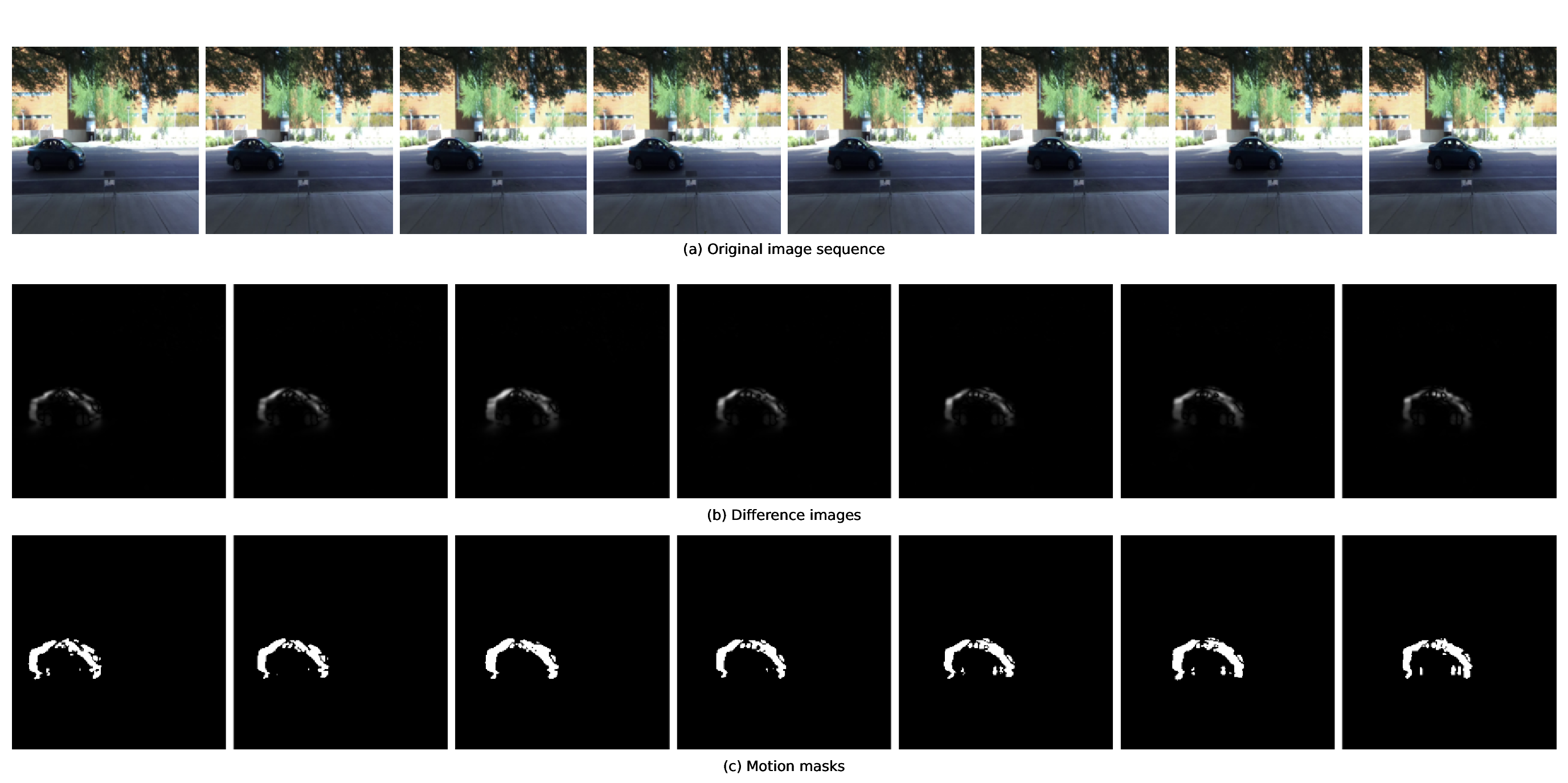}
	\vspace{-2mm}
	\caption{Illustration of data preprocessing with $L=8$. (a): The original sequence of $8$ images. (b): The sequence of $7$ difference images (c): The sequence of $7$ motion masks. }
	\label{fig:preprocessing}
	\vspace{-4mm}
\end{figure*}
The ML models are implemented using PyTorch and trained on NVIDIA Tesla V100 GPUs. In the training stage, we set the batch size to $B=32$ and the maximum number of epochs to $E=100$. The initial learning rate is $10^{-4}$ and a cyclic cosine annealing scheduler is used. Based on preliminary empirical tuning on the validation set to ensure stable convergence and effective class reweighting, we set $\alpha=1$ and $\gamma=2$ in the focal loss. To prevent overfitting, we use a weight decay of $10^{-4}$ and clip the gradient to be no more than $10$. Furthermore, an early stopping technique is adopted to improve training efficiency; in particular, the training process is terminated once the validation loss stops decreasing over $20$ consecutive epochs or the maximum number of epochs is reached. While alternative KD strategies such as feature-based and relational distillation exist, our empirical studies show that logits-based KD achieves superior performance. This is likely due to the architectural mismatch and significant capacity gap between the teacher and student models.

% \blue{The hidden size of the GRUs and the feature size are set to $64$ for both the teacher and student models. However, the teacher model uses a two-layer GRU network, while only a single-layer GRU network is employed in the student model.} In both models, we use $I=8$ heads for diverse attentions. As a result, the teacher model has approximately $1.8 \times 10^6$ total trainable parameters, compared to only $1.7\times 10^5$ trainable parameters for the student model, representing a complexity reduction of over $90\%$.

Fig.~\ref{fig:preprocessing} illustrates the data preprocessing operation $g(\cdot)$ with $L=8$. Fig.~\ref{fig:preprocessing}(a)--(c) shows a sequence of $8$ raw images, and the $7$ corresponding difference and motion masks, respectively. It can be seen that the motion masks highlight the moving UE while effectively removing the background noise.

\subsection{Performance Metrics}
The task loss at the inference stage reflects the overall generalization performance. Besides the task loss, we adopt the Top-$K$ accuracy to measure whether the ground-truth label is among the model's Top-$K$ predicted labels. For the $M=774$ validation samples, let $\hat{y}_{m,1}, \hat{y}_{m,2}, \dots, \hat{y}_{m,K}$ be the Top-$K$ predicted classes (e.g., highest $K$ logits) for sample $m$. The Top-$K$ accuracy is then defined as
\begin{equation}
    \text{Top-}K\text{~Accuracy} = \frac{1}{M} \sum_{m=1}^{M} \mathbf{1} \left( y_m \in \{\hat{y}_{m,1}, \dots, \hat{y}_{m,K} \} \right),
\end{equation}
where $y_m$ denotes the ground-truth label, and $\mathbf{1}(\cdot)$ is the indicator function ($1$ if true, $0$ otherwise). The Top-$K$ score is based on ``hard'' decisions, which may be unnecessary in practice. Given that the close-to-optimal beams may be sufficient for guaranteeing the desired quality of service, a distance-based accuracy (DBA) metric is introduced for the beam prediction task \cite{charan2022multi}, which assigns a score based on the distance between the predicted and ground-truth beams. Specifically, based on the Top-$3$ predicted beams, the DBA score is given by 
\begin{equation}
    \text{DBA}=\frac{1}{3}\sum_{k=1}^3 Y_k
\end{equation}
where
\begin{equation}
    Y_k=1-\frac{1}{M} \sum_{m=1}^M \min _{1 \leq i \leq k} \min \left(\frac{\left|\hat{y}_{m, i}-y_m\right|}{\Delta}, 1\right),
\end{equation}
and $\Delta$ is a normalization factor determining the maximum tolerable distance between the optimal and predicted beams. The term $ \min\left(\frac{|\hat{y}_{m,i} - y_m|}{\Delta}, 1\right)$ serves as a normalized penalty. A small $\Delta$ will induce a large penalty near $1$, making the DBA more sensitive to prediction errors. On the contrary, a large $ \Delta $ indicates tolerance for larger deviations before the penalty is capped at $1$. Unless otherwise mentioned, we set $\Delta=5$ for performance evaluation \cite{charan2022multi}. 

Note that both the Top-$K$ and DBA scores target only one time slot. To reflect the overall performance across all $J+1$ time slots, we further define the average Top-$K$ (ATop-$K$) and average DBA (ADBA) over all time slots, given by
\begin{align}
& \text{ATop-}K\text{~Accuracy}= \frac{1}{J+1}\sum_{\tau=t}^{t+J} \text{Top-}K[\tau], \\
    & \text{ ADBA}= \frac{1}{J+1}\sum_{\tau=t}^{t+J}{\text{DBA}}[\tau],
\end{align}
where $\text{Top-}K[\tau]$ and $\text{DBA}[\tau]$ represent the Top-$K$ accuracy and DBA score at time slot $\tau$, respectively.

\subsection{Overall Performance}\label{sec:overall performance}

\begin{table}[t]
% \small
\centering
\caption{Overall generalization performance of the teacher model in $\%$ for ATop-$k$ accuracy and ADBA score.}\label{tb:teacher model}
\begin{tabular}{r|c|c|c|c}
\hline
Metric &W/o MHA &   With MHA         & Self-KD & YOLO-aided \cite{jiang2022computer} \\
\hline
Test loss & $1.141$    &   $1.050$           &     $\bf 1.016$          &     $0.8158$  \\
\hline
 ATop-$1$ & $40.77$  &     $42.91$         &     $\bf 44.94$          &     $50.20$        \\
 \hline
 ATop-$3$ & $77.44$  &     $79.97$         &     $\bf 81.45$          &     $ 87.15$        \\
 \hline
  ATop-$5$ & $92.31$  &     $93.35$         &     $\bf 94.63$          &     $96.81$        \\

\hline
 ADBA & $93.50$  &     $94.47$         &     $\bf 95.00$          &     $96.63$        \\

\hline
\end{tabular}
\end{table}

\begin{table}[t]
% \small
\centering
\caption{Overall generalization performance of the student model in $\%$ for ATop-$k$ accuracy and ADBA score.}\label{tb:student model}
\begin{tabular}{l|l|ccc}
\hline
Metric& Methods & $L=8$ & $L=5$ & $L=3$ \\
\hline
\multirow{2}{*}{Test loss} & W/o KD & $1.333$ & $1.419$ & $1.471$ \\
& With KD & $1.037$ & $1.046$ & $1.085$ \\
\hline
\multirow{2}{*}{ATop-$1$}  & W/o KD & $38.02$ & $37.28$ & $35.68$ \\
& With  KD & $43.37$ & $43.24$ & $41.79$ \\
\hline
\multirow{2}{*}{ATop-$3$}  & W/o KD & $\bf 73.31$ & $70.62$ & $\bf 68.77$ \\
& With  KD & $\bf 80.25$ & $80.22$ & $\bf 79.25$ \\
\hline
\multirow{2}{*}{ATop-$5$}  & W/o KD & $\bf 88.46$ & $87.45$ & $86.42$ \\
& With  KD & $94.14$ & $94.09$ & $\bf 93.30$ \\
\hline
\multirow{2}{*}{ADBA}  & W/o KD & $\bf 91.04$ & $89.81$ & $88.59$ \\
& With  KD & $94.61$ & $94.29$ & $\bf 94.08$ \\
\hline
\end{tabular}
\end{table}

Table~\ref{tb:teacher model} summarizes the overall generalization performance of the teacher model, where ``W/o MHA'' and ``With MHA'' represent the architecture without and with the MHA module, respectively. In the self-KD scheme, the teacher model contains the MHA module and is trained with guidance from the architecture ``With MHA''. In \cite{jiang2022computer}, Jiang {\it et al.} first obtain the coordinates of potential sensing targets from images with YOLOv4 \cite{bochkovskiy2020yolov4} and manually selects the correct target to be sensed as the input of the neural network consisting of GRUs and MLPs. Since the interference has been removed from the dataset before training, the design in \cite{jiang2022computer} achieves the best generalization performance, and serves as an upper bound. It is seen that the MHA module can effectively enhance the generalization performance of the teacher, increasing the accuracies of ATop-$1$ and ATop-$3$ by 2 percentage points. Moreover, with self-KD, the performance of the teacher is significantly improved. For instance, $95\%$ ADBA is achieved with self-KD, which is over $98\%$ of the optimal accuracy. Note that approximately $6.4\times 10^7$ model parameters are required in Jiang's design (including YOLOv4). In contrast, the teacher model with MHA has only approximately $1.8\times 10^6$ parameters, achieving a reduction in complexity of over $97\%$.

Table~\ref{tb:student model} shows the overall generalization performance of the student model with $L=8,5,3$. We draw the following observations. First, the generalization performance of the student model generally deteriorates with shorter input sequences. For example, the ATop-$3$ accuracy of the student without KD (with KD) is reduced by $5$ ($1$) percentage points when $L$ decreases from $8$ to $3$. This is because a shorter observation window limits the available motion context for inferring future beam evolution, rendering both temporal continuity and accumulated spatial–temporal cues less informative and thereby increasing uncertainty in long-horizon prediction. Knowledge transferred from the teacher can partially compensate for this information loss, as evidenced by the significant performance improvement of the student model with KD. For example, the student model with $L=3$ outperforms the ``vanilla'' student trained from the dataset without KD for $L=8$. In particular, the former attains $94.08\%$ ADBA and $93.30\%$ ATop-$5$ accuracy, while the corresponding values for the latter are $91.04\%$ and $88.46\%$, respectively. Second, with the aid of KD, the student model can achieve performance comparable to the vanilla teacher model without self-KD. For example, the student model trained with KD for $L=8$ achieves $94.61\%$ ADBA, compared with $94.47\%$ for the teacher model without KD. Moreover, the student model trained with KD for $L=3$ achieves $99\%$ the performance of the self-KD teacher, despite the fact that the former requires only $37.5\%$ of the input data and $16.7\times$  fewer parameters than the latter, significantly improving data utilization efficiency and reducing model complexity. 

We observe that the training behavior and final performance are sensitive to the KD hyperparameters, since $\beta$ balances hard-label supervision and the teacher's soft-label supervision, while $\Gamma$ controls the smoothness of the teacher distribution. Therefore, a validation-based grid search is adopted for determining their values. Specifically, we search over the sets $\beta \in \{0.1,0.2,0.3,0.4,0.5\}$ and $\Gamma \in \{2,3,4,5\}$. For each candidate pair, the considered model is trained using the same training protocol, and the checkpoint with the lowest validation loss is retained. The final values of $\beta$ and $\Gamma$ are chosen based on the best validation performance among all candidate pairs. Table~\ref{tb:beta-temperature} summarizes the optimal values of $\beta$ and $\Gamma$. It can be observed that the student model benefits from a larger temperature compared to the teacher self-distillation setting. This is because the teacher model is already well trained using ground-truth supervision, making the additional soft-label information from KD less influential. In contrast, the lower-capacity student relies more on softened teacher targets to effectively transfer knowledge, leading to a preference for higher temperature values.

\begin{table}[t]
% \small
\centering
\caption{Values of $\beta$ and $\Gamma$ for KD-aided learning.}\label{tb:beta-temperature}
\begin{tabular}{r|c|c|c|c}
\hline
\multirow{2}{*}{Term} & \multirow{2}{*}{Teacher (self-KD)} & \multicolumn{3}{c}{Student} \\
\cline{3-5}
 &  & $L=8$ & $L=5$ & $L=3$ \\
\hline
$\beta$ & $0.3$ & $0.4$ & $0.2$ & $0.4$ \\
\hline
$\Gamma$ & $2$ & $3$ & $5$ & $4$ \\
\hline
\end{tabular}
\end{table}

\begin{figure*}[t]
\vspace{-6mm}
\small
    \centering
    %\hspace{-5mm}
        \subfigure[Top-$3$ and Top-$5$ prediction accuracy.]
    {\label{fig:Top3n5_Acc_L8} \includegraphics[width=0.45\textwidth]{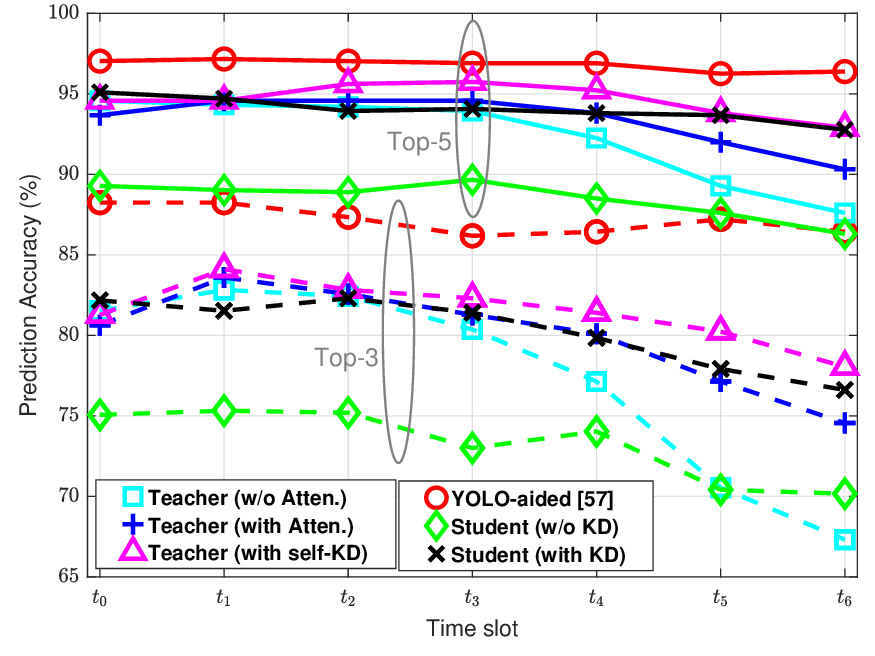}}%\hspace{-3mm}
     % \hspace{-10mm}
        \subfigure[DBA score.]
    {\label{fig:DBA_L8} \includegraphics[width=0.45\textwidth]{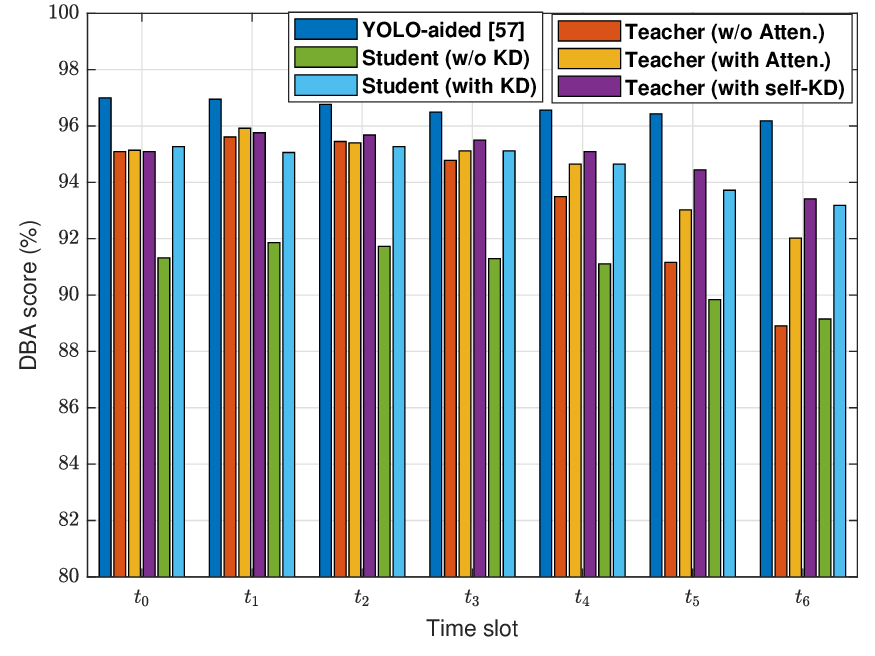}}%\hspace{-3mm}
    \vspace{-2mm}
    \caption{Performance of the teacher and student models with $L=8$.}
    \label{fig:performance L8}
    \vspace{-5mm}
\end{figure*}

\begin{figure*}[t]
% \vspace{-2mm}
\small
    \centering
    %\hspace{-5mm}
        \subfigure[Top-$3$ and Top-$5$ prediction accuracy.]
    {\label{fig:Top3n5_Acc_L8_cmp} \includegraphics[width=0.45\textwidth]{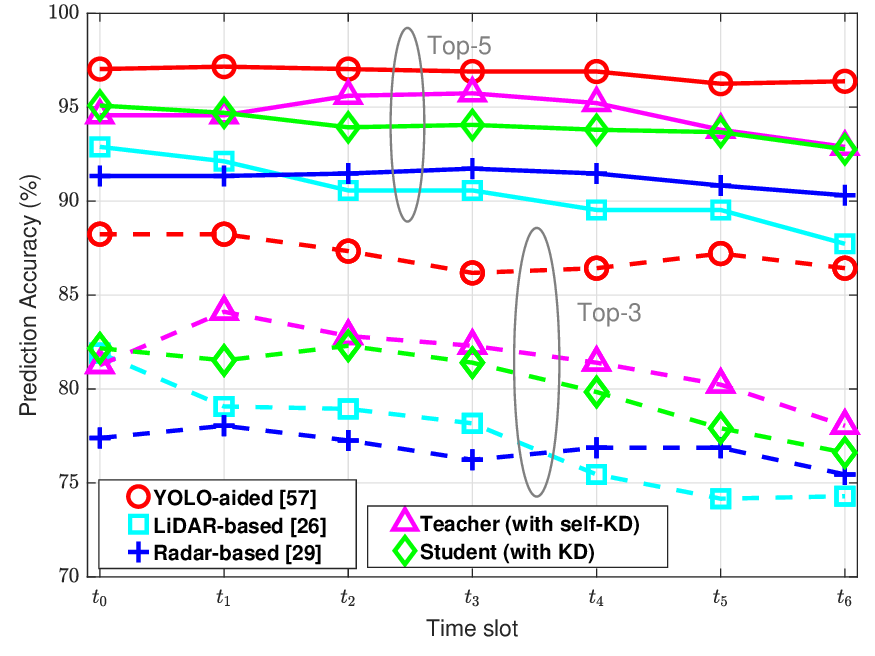}}%\hspace{-3mm}
     % \hspace{-10mm}
        \subfigure[DBA score.]
    {\label{fig:DBA_cmp} \includegraphics[width=0.45\textwidth]{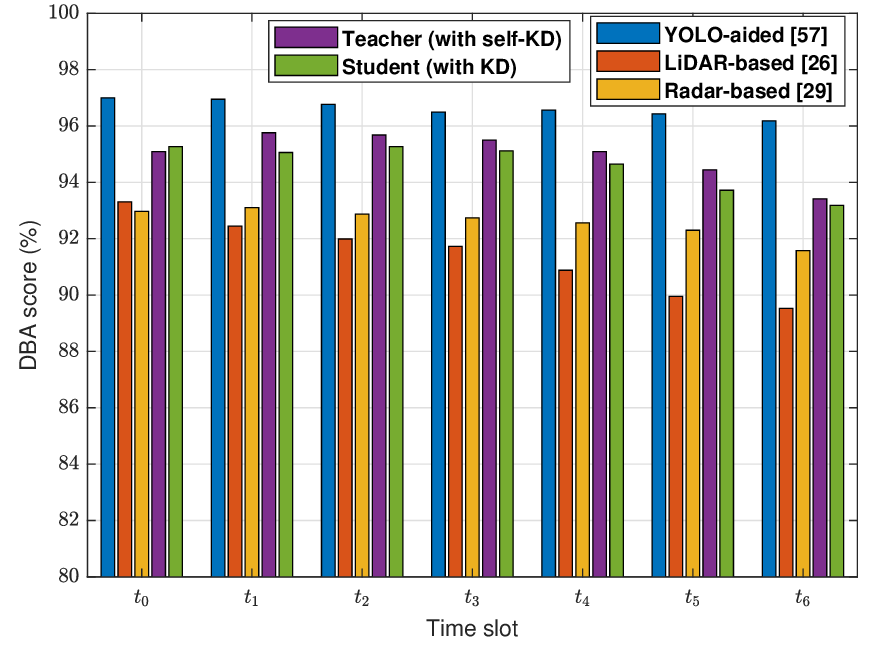}}%\hspace{-3mm}
    \vspace{-2mm}
    \caption{Performance comparison  with different modalities ($L=8$).}
    \label{fig:modality comparison}
    \vspace{-2mm}
\end{figure*}

\begin{table*}[!h]
\centering
\caption{Model complexity and latency comparison under different sensing modalities and lengths $L$ of input sequence.}
\label{tab:complexity_latency}
\begin{tabular}{l|l|l|l|l}
\hline
\textbf{Category} & \textbf{Model} & \textbf{Params (M)} & \textbf{FLOPs (M)} & \textbf{Latency (mean$\pm$std~ms)} \\
\hline
\hline
\multirow{4}{*}{Image} 
 & Teacher ($L=8$)  &$ 1.788$ & $156.474$ & $ 9.989 \pm 2.145 $ \\
 \cline{2-5}
&Student ($L=8$) & $ 0.107 $ ($16.7\times$ fewer)& $92.216$ ($1.7\times$ fewer)& $9.410\pm 1.392$ \\
&Student ($L=5$) & $ 0.107 $ & $57.635$ ($2.7\times$ fewer)& $ 7.384 \pm  1.075$ \\
&Student ($L=3$) & $ 0.107 $ & $34.581$ ($4.5\times$ fewer)& $ 6.019 \pm 0.853$ ($1.6\times$ fewer) \\
\hline
Radar 
 & \cite{Luo2023millimeter} ($L=8$)   & $0.275$ & $404.752 $ & $19.448 \pm 3.171 $ \\
\hline
LiDAR 
 & \cite{Jiang2024LiDAR} ($L=8$)   & $0.043$ & $13.152$ & $0.250 \pm 0.029  
$ \\
\hline
\end{tabular}
\vspace{-4mm}
\end{table*}

\begin{figure*}[t]
% \vspace{-2mm}
\small
    \centering
    % \hspace{-5mm}
     % \hspace{-10mm}
    \subfigure[Student model with $L=5$.]
        {\label{fig:Top3n5_Acc_L5} \includegraphics[width=0.45\textwidth]{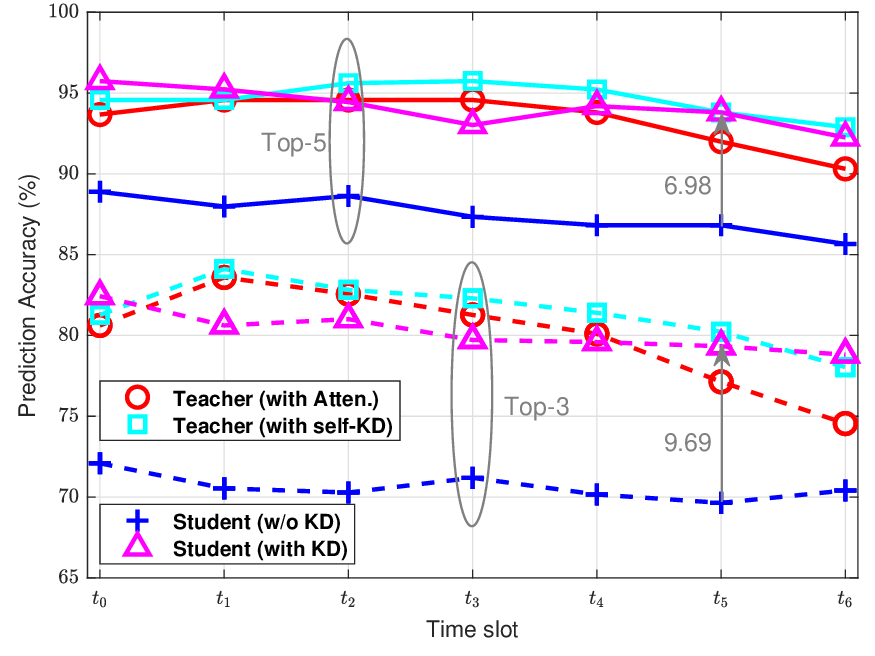}}%\hspace{-3mm}
    \subfigure[Student model with $L=3$.]
        {\label{fig:Top3n5_Acc_L3} \includegraphics[width=0.45\textwidth]{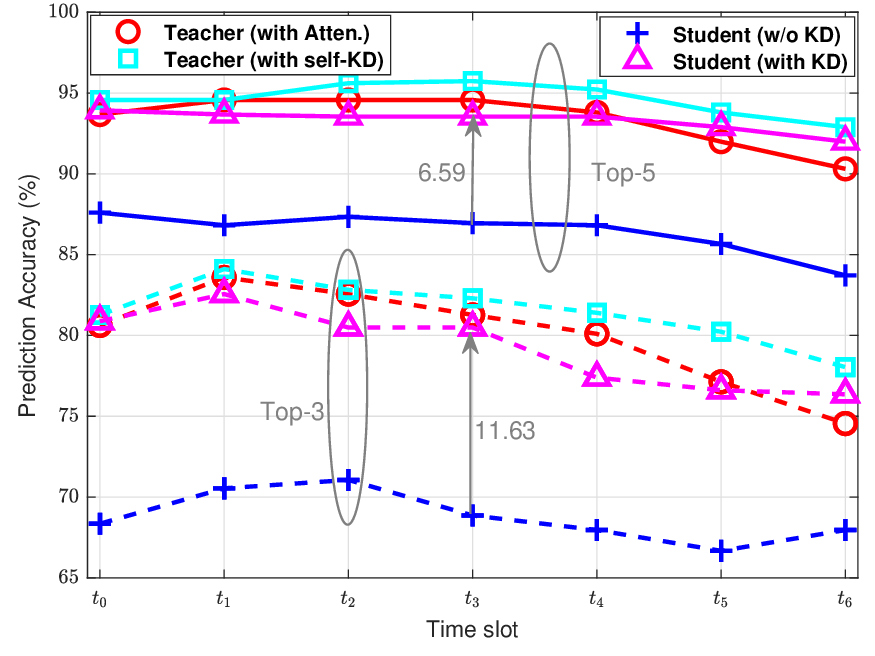}}%\hspace{-3mm}
    \vspace{-2mm}
    \caption{Top-$3$ and Top-$5$ prediction accuracy of the student model. The teacher models are trained and tested under $L=8$.}
    \label{fig:prediction performance L35}
    \vspace{-2mm}
\end{figure*}

\begin{figure*}[t]
\vspace{-2mm}
\small
    \centering
    % \hspace{-5mm}
    \subfigure[Student model with $L=5$.]
        {\label{fig:DBA_L5} \includegraphics[width=0.45\textwidth]{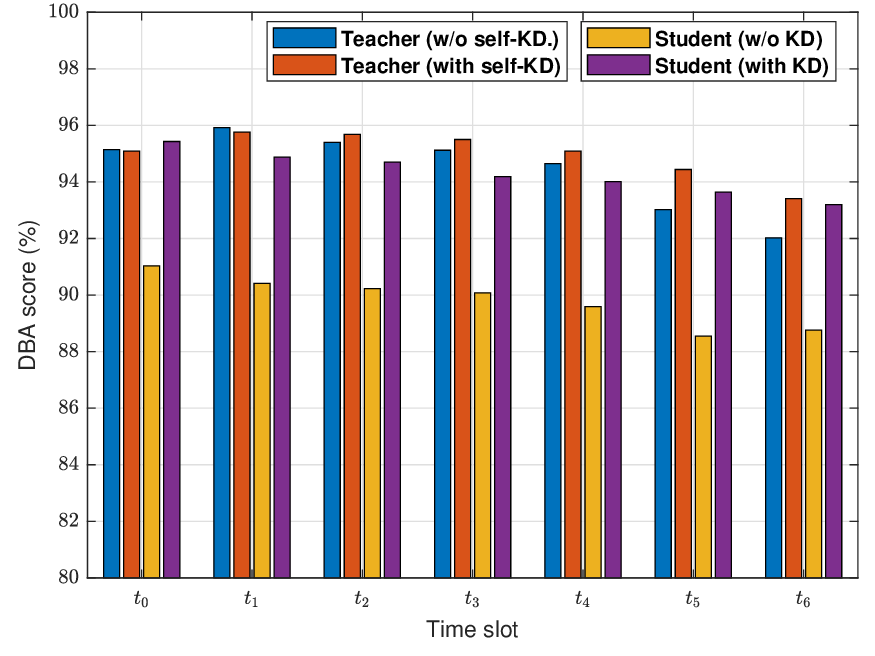}}%\hspace{-3mm}
    \subfigure[Student model with $L=3$.]
        {\label{fig:DBA_L3} \includegraphics[width=0.45\textwidth]{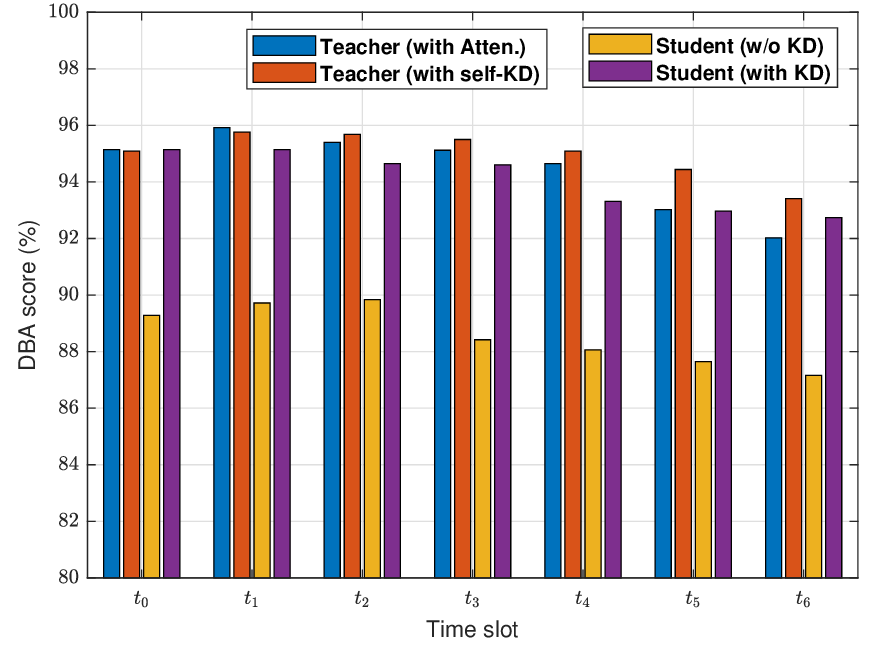}}%\hspace{-3mm}
    \vspace{-2mm}
    \caption{DBA score of the student model with $L=3,5$. The teacher models are trained and tested under $L=8$.}
    \label{fig:Comparision of DBA score versus L}
    \vspace{-5mm}
\end{figure*}

\subsection{Performance for Specific Time Slots}\label{sec: detailed performance}
Fig.~\ref{fig:performance L8} shows the performance of the teacher and student models with $L=8$. It is seen that, as expected, the teacher and student achieve lower prediction accuracy and DBA scores for beam prediction farther into the future. %This is reasonable since the beam prediction becomes more challenging for the far future.
Comparing the performance of the teacher with and without MHA in Fig.~\ref{fig:Top3n5_Acc_L8}, we observe that the integrated attention mechanism primarily improves the prediction accuracy of later time slots, i.e., $t_4$--$t_6$. Such results verify that the MHA can effectively capture time dependencies inherent in the input sequence, extracting useful information for more challenging future beam predictions. Similarly, the teacher with self-KD mainly improves the performance of beam prediction for time slots farther in the future. With self-KD, the additional soft distribution information between candidate beam classes provides an extra regularizer that helps the model learn better for more challenging tasks \cite{mobahi2020self}. The effectiveness of KD is particularly prominent for the student model, which has significant gaps in prediction accuracy compared to the teacher when implemented without KD-aided learning. In contrast, with KD, the student model achieves close to $95\%$ Top-$5$ accuracy, which even surpasses the vanilla teacher without self-KD at the future time slots $t_5$ and $t_6$. Although the Top-$3$ accuracy of the teacher and student models is no more than $85\%$, the corresponding DBA scores can reach $95\%$, as seen in Fig.~\ref{fig:DBA_L8}. 

Fig.~\ref{fig:modality comparison} compares the prediction performance of the proposed image-based models with state-of-the-art benchmarks, including the YOLO-aided image-based solution in \cite{jiang2022computer}, the LiDAR-based design of \cite{Jiang2024LiDAR}, and the radar-based approach in \cite{Luo2023millimeter}. All methods use a sequence length $L=8$. As shown in Fig.~\ref{fig:Top3n5_Acc_L8_cmp}, both the proposed teacher and student models outperform the LiDAR- and radar-based schemes, particularly for time slots farther in the future. For example, at future time slot $t_6$, the student model achieves Top-5 accuracy gains of $2.45$ and $5.03$ percentage points over the radar- and LiDAR-based methods, respectively. This advantage is further confirmed by the DBA score comparison in Fig.~\ref{fig:DBA_cmp}. We further compare model size, computational complexity, and inference latency in Table~\ref{tab:complexity_latency}, where the latency is measured with batch size 1 using CUDA events over $300$ iterations (after $100$ warm-up iterations), as reported on a Tesla-V100 platform. The radar-based approach exhibits the highest FLOP count and latency, whereas the LiDAR-based method has the lowest computational cost among the considered methods. Although the teacher model achieves the best prediction accuracy, it has substantially more parameters. In contrast, the student model reduces model size by over $16.7\times$ and the FLOP count by over $1.7\times$ relative to the teacher. Since computational cost scales with input sequence length, the student model with $L=3$ further reduces the FLOP count and latency by $4.5\times$ and $1.6\times$, respectively, compared with the teacher. Overall, the student model provides a favorable trade-off between prediction accuracy and computational efficiency among the considered approaches.

Figs.~\ref{fig:Top3n5_Acc_L5} and \ref{fig:Top3n5_Acc_L3} show the prediction accuracy of the student models with $L=5$ and $L=3$, respectively, where the teacher is trained and tested with $L=8$ for comparison. The teacher model with/without self-KD specifically refers to the architecture employing MHA. It is observed that KD significantly improves the performance of the student model, especially for Top-$3$ prediction. For example, percentage-point enhancements of $9.69$ and $6.98$ in Top-$3$ and Top-$5$ accuracy are achieved at time slot $t_6$ for KD-aided learning for $L=5$. A similar trend is observed for $L=3$, as shown in Fig.~\ref{fig:Top3n5_Acc_L3}. This is because Top-3 prediction is inherently more challenging than Top-5 prediction, making the additional knowledge provided by the teacher more beneficial. Furthermore, the student with KD for both $L=5$ and $L=3$ achieves slightly higher Top-$5$ prediction accuracy at $t_5$ and $t_6$ than the vanilla teacher model with $L=8$. The student with KD achieves over $92\%$ Top-$5$ prediction accuracy for both $L=5$ and $L=3$, showcasing the effectiveness of KD-aided learning. It is also verified from Figs.~\ref{fig:DBA_L5}~and~\ref{fig:DBA_L3} that the DBA scores, which are based on the Top-$3$ accuracy, can reach as high as $95\%$ for both $L=5$ and $L=3$.

\begin{figure}[t]
\vspace{-6mm}
\small
    \centering
    % \hspace{-4mm}
    %     \subfigure[Top-$1$ accuracy.]
    % {\label{fig:Top1_improvement} \includegraphics[width=0.5\textwidth]{Figs_R1/Top1_improvement.eps}}%\hspace{-2mm}
     % \hspace{-10mm}
    \subfigure[Top-$3$ accuracy.]
        {\label{fig:Top3_improvement} \includegraphics[width=0.45\textwidth]{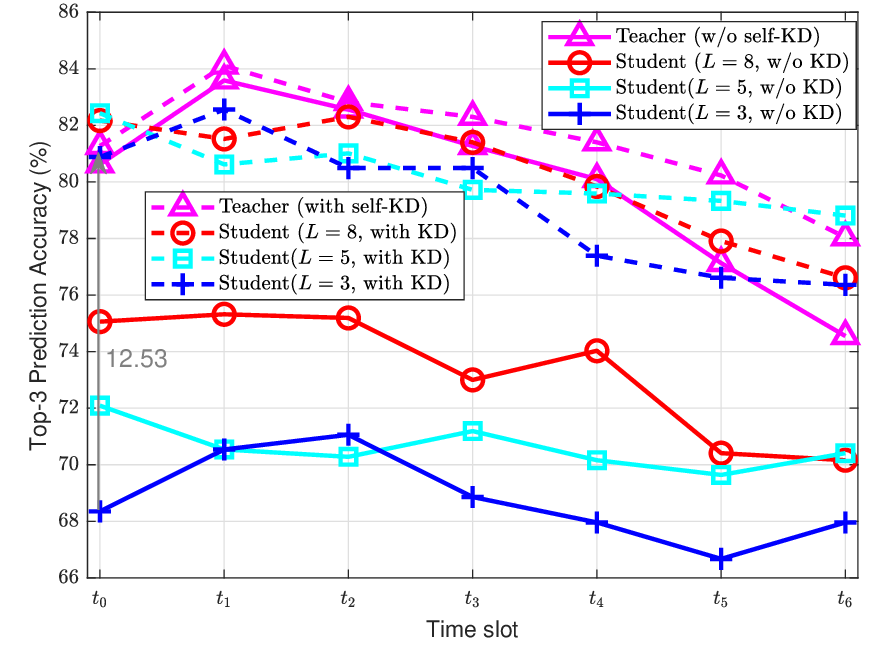}}
        \hspace{-2mm}
    \subfigure[Top-$5$ accuracy.]
        {\label{fig:Top5_improvement} \includegraphics[width=0.45\textwidth]{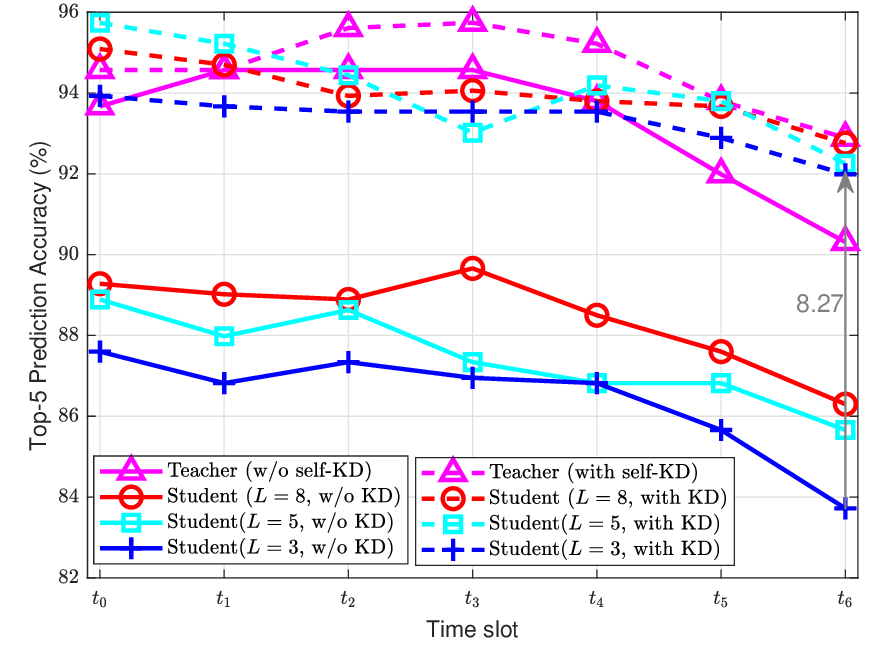}}
        \hspace{-2mm}
    \subfigure[DBA score.]
        {\label{fig:DBA_improvement} \includegraphics[width=0.45\textwidth]{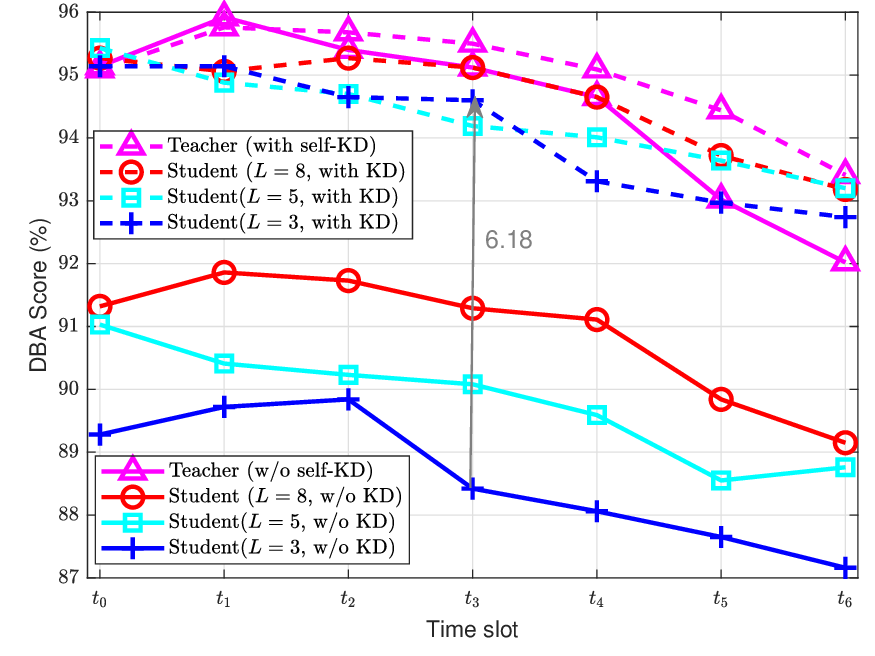}}%\hspace{-2mm}
    \vspace{-2mm}
    \caption{Performance of the teacher and student models.}
    \label{fig:Performance of the teacher and student models.}
    % \vspace{-2mm}
\end{figure}

Figs~\ref{fig:Top3_improvement}--\ref{fig:DBA_improvement} show the Top-$3$ and Top-$5$ prediction accuracies and DBA scores of the teacher and student models, respectively. Several important observations can be made. First, KD consistently improves the prediction performance of both the teacher and student models across all metrics and time slots, confirming its effectiveness for beam prediction. KD provides maximum gains of approximately $12.53$, $8.27$, and $6.18$ percentage points in Top-$3$, Top-$5$, and DBA, respectively, for the student model with $L=3$. Second, with KD, the student model approaches the teacher performance, especially for Top-$5$ accuracy and DBA score, demonstrating effective knowledge transfer from the high-capacity teacher to the lightweight student. For example, the KD-aided student model with $L=3$ can surpass the vanilla teacher model at time slots $t_5$ and $t_6$ farther in the future, achieving an enhancement of $1.68$ Top-$5$ percentage points compared to the vanilla teacher model at $t_6$. Third, prediction accuracy gradually decreases for more distant future time slots due to the increased uncertainty in long-term beam evolution. Consequently, longer input sequences yield better performance %($L=8 \geq L=5 \geq L=3$), 
since richer temporal context facilitates more accurate beam trajectory inference.

 \begin{figure}[tb]
\vspace{-6mm}
	\small
		% \vspace{-4mm}
	\centering	
	\includegraphics[width=0.45\textwidth]{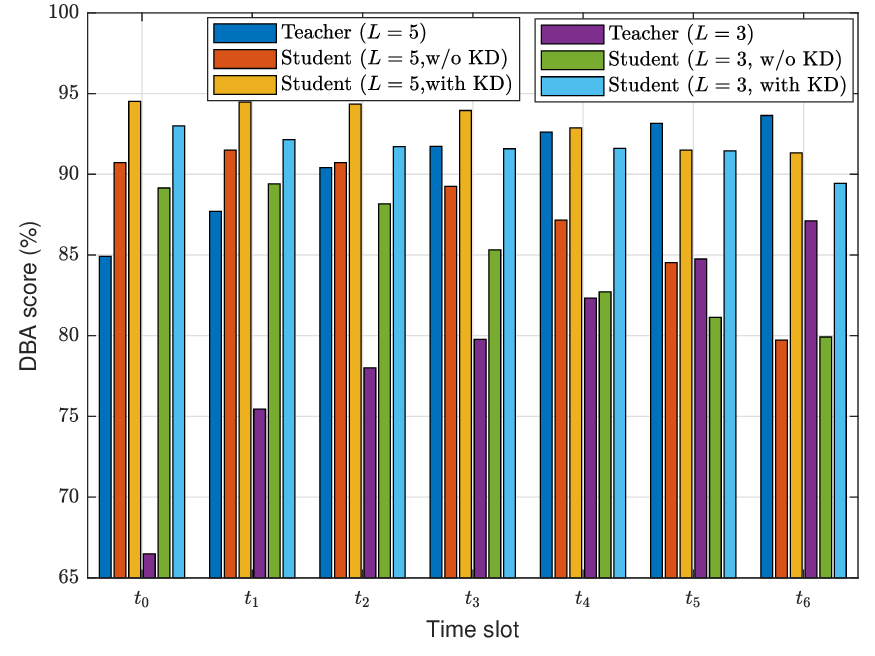}
	\vspace{-2mm}
	\caption{Generalization performance of the teacher model (with self-KD). The teacher model is trained for $L=8$ and tested for $L=5,3$. In contrast, the student models for comparison are trained and tested for $L=5$ and $L=3$, respectively.}
	\label{fig:teacher generilization}
	% \vspace{-2mm}
\end{figure}

Fig.~\ref{fig:teacher generilization} demonstrates the generalization performance of the self-KD-aided teacher model, which is trained for $L=8$ and tested for $L=5,3$. In contrast, the student models for comparison are trained and tested for $L=5$ and $L=3$, respectively. It is seen that the DBA score achieved by the teacher is lowest at time slot $t_0$ and highest at time slot $t_6$. The surprising increase in accuracy for farther time slots with shorter inputs likely arises from a training-testing mismatch, where the model struggles early due to reduced input context but benefits from internal GRU decoder dynamics (e.g., autoregressive accumulation of the hidden states) in father time slots. In contrast, the student with KD maintains relatively high DBA scores across all time slots, verifying its robustness.

\subsection{Ablation Study on Codebook Size and Scenario Variations}\label{sec:ablation}

\begin{figure*}[t]
\vspace{-6mm}
\small
    \centering
    % \vspace{-5mm}
        \subfigure[Top-$3$ and Top-$5$ prediction accuracy.]
    {\label{fig:Top3n5_C64C32} \includegraphics[width=0.45\textwidth]{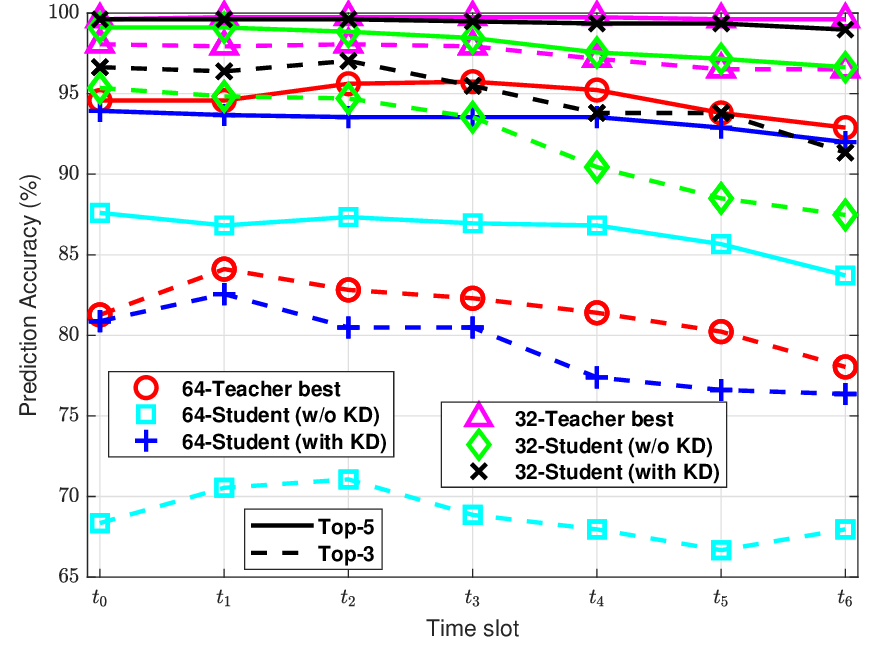}}%\hspace{-3mm}
     % \hspace{-10mm}
        \subfigure[DBA score.]
    {\label{fig:DBA_C64C32} \includegraphics[width=0.45\textwidth]{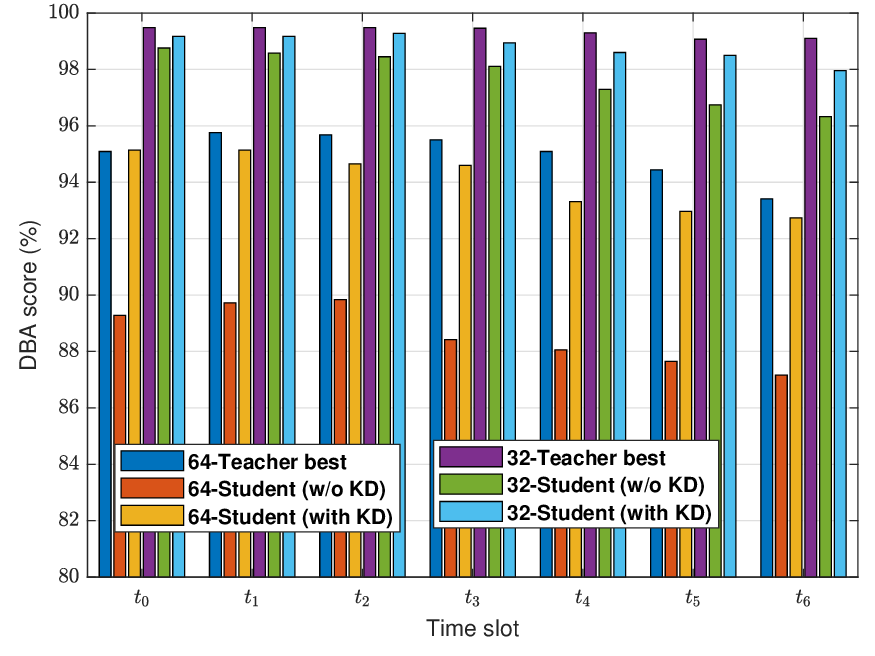}}%\hspace{-3mm}
    \vspace{-2mm}
    \caption{Performance comparison with for $C=32$ and $C=64$.}
    \label{fig:performance C64C32}
    % \vspace{-2mm}
\end{figure*}

\begin{figure*}[t]
\vspace{-4mm}
\small
    \centering
    %\hspace{-5mm}
        \subfigure[Top-$3$ and Top-$5$ prediction accuracy.]
    {\label{fig:Top5n3_S8S9} \includegraphics[width=0.45\textwidth]{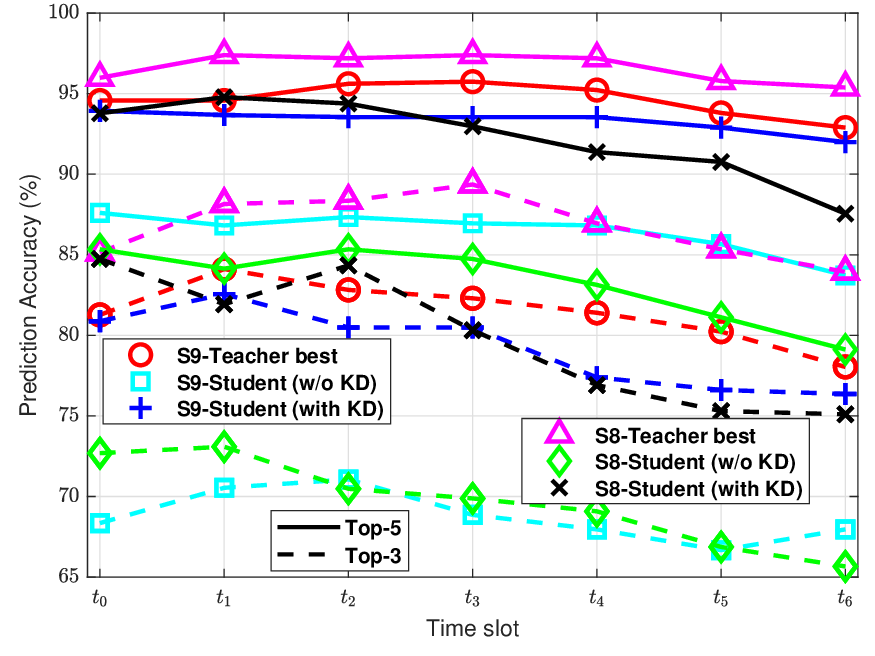}}%\hspace{-3mm}
     % \hspace{-10mm}
        \subfigure[DBA score.]
    {\label{fig:DBA_S8S9} \includegraphics[width=0.45\textwidth]{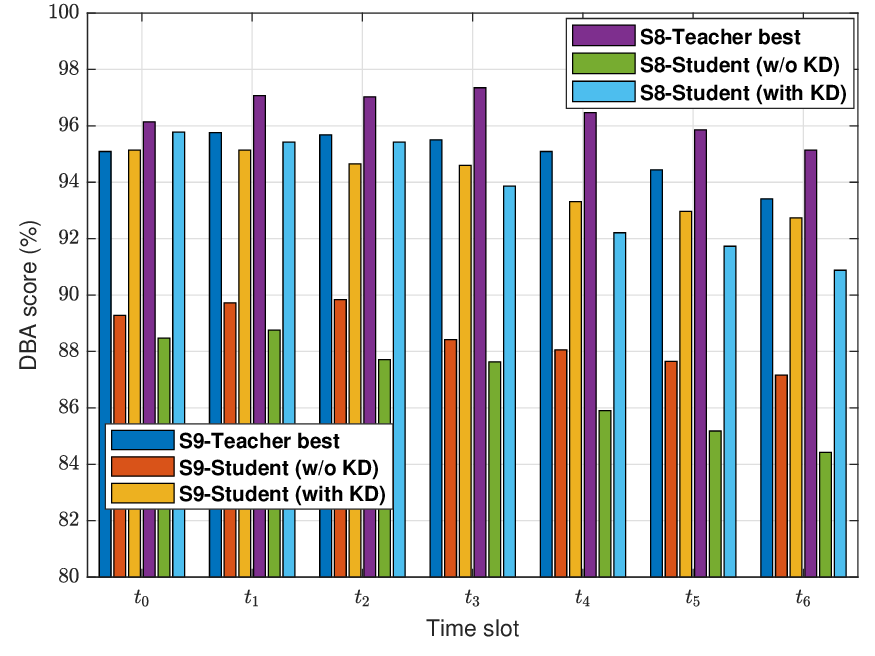}}%\hspace{-3mm}
    \vspace{-2mm}
    \caption{Performance evaluation based on scenarios 8 and 9 of DeepSense 6G dataset.}
    \label{fig:performance S8S9}
    \vspace{-6mm}
\end{figure*}

Fig.~\ref{fig:performance C64C32} shows the performance of the proposed designs under two beam codebook sizes, $C=64$ and $C=32$, denoted by the prefixes “$64$-” and “$32$-” in the labels. In the figures, ''Teacher best'' represents the teacher model with the best performance, while ''Student'' refers to the student model with $L=3$. For $C=32$, the teacher and student architectures are identical to those for $C=64$, except for the reduced output dimension of the classifier. We observe that reducing the codebook size consistently improves Top-$3$/Top-$5$ accuracy and DBA scores for both the teacher and student models, since beam selection becomes less ambiguous with fewer candidate beams. As shown in Fig.~\ref{fig:Top3n5_C64C32}, the teacher achieves over $96\%$ Top-$3$ accuracy and over $99\%$ Top-5 accuracy for $C=32$. Moreover, both teacher and KD-aided student models attain DBA scores above $98\%$ across all time slots, whereas the corresponding values for $C=64$ remain below $96\%$, as shown in Fig.~\ref{fig:DBA_C64C32}. Notably, the relative performance ranking among the methods remains unchanged across codebook sizes, i.e., Teacher $>$ Student (with KD) $>$ Student (without KD), confirming the effectiveness of KD and demonstrating the robustness of the proposed lightweight design to different beam resolutions.

Fig.~\ref{fig:performance S8S9} presents the evaluated model performance for $C=64$ on Scenarios 8 and 9 of the DeepSense 6G dataset, denoted by the prefixes “S8-” and “S9-” in the labels. In the figures, “Teacher best” denotes the teacher model with the best performance, while “Student” refers to the student model with $L=3$. Scenario 8 contains $4043$ samples, with $80\%$ used for training and $20\%$ for testing. Scenarios 8 and 9 share the same physical location and sensing testbed but exhibit different traffic patterns. The teacher and student architectures used for Scenario~8 are identical to those for Scenario~9. From Figs.~\ref{fig:Top5n3_S8S9}~and~\ref{fig:DBA_S8S9}, we observe a consistent performance ranking in both scenarios, i.e., Teacher $>$ Student (with KD) $>$ Student (without KD), indicating that the proposed architecture and KD framework remain stable across varying traffic conditions. Moreover, KD provides substantial performance gains for the student model in both cases, confirming robust knowledge transfer under scene variation. These results demonstrate that the proposed framework maintains effectiveness across different trajectory and traffic conditions within the same deployment settings.

%  \begin{figure}[t]
% % \vspace{-3mm}
% 	\small
% 		% \vspace{-4mm}
% 	\centering	
% 	\includegraphics[width=0.5\textwidth]{Figs/testloss_std_seqlen.eps}
% 	\vspace{-2mm}
% 	\caption{Test performance of the student model versus input sequence length $L$ without KD.}
% 	\label{fig:std_testloss_noKD}
% 	% \vspace{-4mm}
% \end{figure}

% \vspace{5mm}

\section{Conclusions}\label{sec:conclusion}
This work has proposed a KD-based vision-assisted long-term beam tracking framework that predicts both current and future beams from past sensor observations. A high-capacity teacher model, built with CNNs, GRUs, and MHA, is first developed to achieve high predictive accuracy. Leveraging KD, a lightweight student model is then trained not only to reduce model size and complexity but also to operate effectively with shorter input sequences while preserving long-term prediction ability. Experimental results show that the teacher approaches state-of-the-art long-term prediction accuracies with a $90\%$ reduction in complexity. The student model closely matches the teacher’s performance  and maintains over $90\%$ long-term beam prediction accuracy even with $1670\%$ fewer parameters, $450\%$ lower complexity, and $60\%$ shorter input sequences. These advantages substantially reduce power consumption, sensing and processing latency, and signaling overhead, advancing practical deployment in resource-constrained ISAC systems. The proposed approach provides a viable pathway toward high-accuracy, low-latency, and energy-efficient long-term beam tracking. Future research may explore advanced motion representations (e.g., optical flow or detection-based cues) to further improve the proposed framework and extend it to more challenging real-world scenarios involving LoS blockages and cross-geometry domain shifts.

% incorporate multiple sensing modalities for more challenging real-world scenarios

% \FloatBarrier
\bibliographystyle{IEEEtran}
\bibliography{conf_short,jour_short,refs-my}

\end{document}